\documentclass[12pt,twoside,english,a4paper]{article}
\usepackage{caption}
\usepackage[utf8]{inputenc}
\usepackage[intlimits] {amsmath}   % muss for defjs# stehen
\usepackage{graphicx}
\usepackage{psfrag}
\usepackage{ifthen}
\usepackage{fancyhdr}
\usepackage{rotating}
\usepackage{multirow}
\usepackage{booktabs}              % Dicke Tabellenlinien
\usepackage{amssymb}
\usepackage{amsbsy}
\usepackage{bm}                    % fette Schrift in Formeln
\usepackage{bbm}
\usepackage{babel}
\usepackage{theorem}
\usepackage{upgreek}  % gerade (roman) griechische Buchstaben z.B. \upalpha
\usepackage{pifont}   % zusaetzliche Schriften, z.B. eingekreiste Zahlen
\usepackage[active]{srcltx}   % aktive suche im xdv

\usepackage{tikz}
\usepackage{pgfplots}
\usepackage{nicefrac}
\usepackage[ruled]{algorithm2e}
\usepackage{appendix}
\usepackage{longtable}
\usepackage{url}
\usepackage{graphicx}
\usepackage[
    colorlinks=false,
    citebordercolor={cyan}
]{hyperref}
\usepackage{xcolor}
\usepackage{bm}
\usepackage{tabularray}
\usepackage{adjustbox}
\usepackage{transparent}

\renewcommand{\vec}[1]{\bm{#1}}
\newcommand{\ten}[1]{\bm{#1}}

\usetikzlibrary{positioning, calc, shapes.misc, external}
\pgfplotsset{compat=1.18}
\usepackage{suetterl}  % Suetterlin Schrift
\newfont{\suetdbl}{suet14 scaled 2000}  % Suetterlin skaliert um Faktor 2,000

\usepackage{yfonts}  % altdeutsche Schrifte und Initialen
\newfont{\gothdbl}{ygoth scaled 2000} 
\newfont{\frakdbl}{yfrak scaled 2000}
\newfont{\swabdbl}{yswab scaled 2000}
\usepackage[normalem]{ulem}  % zusaetzliche Unterstreichungsformen wie z.B.:
\usepackage[textsize=footnotesize,textwidth=1.7cm]{todonotes}
\usepackage[]{defcommandsls}      % eigene Definitionen: Option BoldVarsRoman

\usepackage[numbers,sort]{natbib}
\fancypagestyle{plain}{%
  \fancyhead{}%
  \fancyfoot[c]{\sffamily\thepage}%
}
\makeatletter                           % Leere Fuellseiten
\def\cleardoublepage{\clearpage\if@twoside \ifodd\c@page\else
  \hbox{}
  \vspace*{\fill}
  \thispagestyle{empty}
  \newpage
  \if@twocolumn\hbox{}\newpage\fi\fi\fi}
\makeatother
\begin{document}
\unitlength1.0cm
\frenchspacing

%=== cover page
%\include{covertemplate_ls}

%=== sections ===
%--- section 1
%\include{sec_1/sec}

\thispagestyle{empty}
{\bf \large
\begin{center}
    Empirical Material Sampling and Linearisation
\end{center}
}

\vspace{-3mm}
\begin{center}
{\bf \large
A Simple and Efficient Strain-Space Model Order Reduction Approach for Computational Homogenisation in Large-Deformation Hyperelasticity
}
\end{center}

\vspace{4mm}
\begin{center}
{Erik Faust \& Lisa Scheunemann}
\end{center}

\vspace{4mm}
\begin{center}{Institute of Applied Mechanics (IFAM)}
{RWTH Aachen University, Mies-van-der-Rohe-Str. 1, 52074 Aachen,
Germany}
{\\Chair of Applied Mechanics (LTM)}
{RPTU Kaiserslautern-Landau, Gottlieb-Daimler-Str.,
67663 Kaiserslautern,
Germany}
{\\ \small e-mails: erik.faust@ifam.rwth-aachen.de, lisa.scheunemann@ifam.rwth-aachen.de
    %phone: +49 (0) 241 80 25003,
    %fax: +49 xxx
    }
    \end{center}

\vspace{4mm}
\begin{center}
{\bf \large Abstract}
\bigskip

{\footnotesize
\begin{minipage}{14.5cm}
\noindent

In this article, we propose a simple and efficient hyperreduced strain-space model order reduction (MOR) approach for hyperelastic representative volume elements (RVEs), called Empirical Material Sampling and Linearisation (EMSL). The approach is conceptually motivated by the Empirically Corrected Cluster Cubature (E3C) of Wulfinghoff  and Hauck~\cite{WulHau:2025:EHR}, but also draws on ideas from previous work on incremental variational structure-preserving strain-space model order reduction techniques to achieve rapid evaluations in the online phase.

As in E3C, we group the material domain into regions of similar behaviour, and query the material routine at one reference strain value per region. However, we sample these strains only once per load increment, at empirically estimated expected strain values. We use the reference material tangent and strain modes obtained via the Proper Orthogonal Decomposition (POD) to compute a {linearised estimate} of the stress response in the remainder of the material cluster. In contrast to E3C, which approximately integrates the exact material law, EMSL could therefore be said to exactly integrate an approximation of the material behaviour.
The resulting reduced problem is affine in each load step, allowing for integration over the entire computational domain via operations which can readily be preprocessed in the offline phase. Since a linear equation system is obtained in each load increment, no Newton iterations are required in the online phase.%convergence is guaranteed in one iteration. %Alternatively, EMSL can be applied iteratively as a fixed-point scheme, yielding slightly higher accuracy at somewhat higher computational cost. 

For benchmark comparisons, we pose a variant of two popular reduced cubature schemes in strain space and recall the E3C algorithm proposed by Wulfinghoff et al. %introduce a slightly modified version of E3C. 
On an example hyperelastic RVE problem with a porous geometry, we show that EMSL Pareto-dominates competing strain-space approaches in terms of the tradeoff between accuracy and runtime.

\end{minipage}
}
\end{center}

{\bf Keywords:}
model order reduction, multiscale modeling, computational homogenisation, strain-space, hyperreduction

\section{Introduction}

Model order reduction (MOR) methods which tackle computational homogenisation problems in strain space {may} be classified into two loose families. Approaches such as the Nonuniform Transformation Field Analysis (NTFA)~\cite{MicSuq:2003:ntf}, Potential-Based Reduced Basis MOR (pRBMOR)~\cite{FriLeu:2013:rbh}, or the Self-Consistent Clustering Analysis (SCCA)~\cite{LiuBesLiu:2016:sca,YUKafLiu:2019:SCC} exploit specific physical and structural features of the homogenisation problem to obtain an incremental description in terms of inelastic modes or cluster-level coefficients. Meanwhile, methods like the Finite Element (FE)-Galerkin scheme proposed in~\cite{KunFri:2019:fsh}, the High-Performance Reduction method from~\cite{RasLloHue:2021:hpr}, the cluster-enhanced potential-based scheme from~\cite{RuaJuChe:2024:CPR}, and the Empirically Corrected Cluster Cubature (E3C) by Wulfinghoff et al.~\cite{WulHau:2025:ech,wul:2025:ECC,WulHau:2025:EHR} employ a more general Galerkin framework along with a hyper-reduced cubature of reduced residuals. As originally proposed in~\cite{HerCaiFer:2017:dhn,Her:2020:MMP}, this reduced cubature can be performed on the integration point level, bypassing element-level formulations.
Both families have yielded impressive results {in reducing the prohibitive computational costs (see e.g.~\cite{KlaKohLan:2020:chm}) common in many multiscale modeling applications, without sacrificing too much in the way of accuracy~\cite{MicSuq:2003:ntf,FriLeu:2013:rbh,LiuBesLiu:2016:sca,YUKafLiu:2019:SCC,KunFri:2019:fsh,WulHau:2025:ech,wul:2025:ECC,WulHau:2025:EHR}.}
%in terms of the tradeoff between accuracy and runtime in a range of applications in multiscale modeling~\cite{MicSuq:2003:ntf,FriLeu:2013:rbh,LiuBesLiu:2016:sca,YUKafLiu:2019:SCC,KunFri:2019:fsh,WulHau:2025:ech,wul:2025:ECC,WulHau:2025:EHR}. 
While the former successfully exploit the specific physics of the target problem class, 
the latter are attractive on account of their generality, conceptual simplicity and low intrusiveness. {Stress and strain data can be used for training}, and access to element routines utilised by a high-fidelity Finite Element (FE) solver is not necessary. Only the original material models are required. This means that hyperreduced model reduction schemes can be deployed quite independently from the underlying high-fidelity solver -- in computational homogenisation, for example, strain-based schemes can be deployed with training data coming either from FE or Fast Fourier Transform (FFT) solvers. A more exhaustive review of the literature on MOR approaches to representative volume element (RVE) problems beyond these strain-space families
can, for example, be found in our previous work~\cite{SchFau:2025:MLA,FauSch:2025:HML}.

{Beyond the articles mentioned above, which motivate our method, the interested reader may also consult~\cite{NasOsk:2026:SEH}, who recently performed reduced integration of the strain and stress fields using linear and higher-order spatial modes in the context of NTFA. Prior to E3C, and in displacement space,~\cite{BenYvoBar:2022:mcm}~also clustered integration points to reduce integration effort, while assuming constant strain in each cluster. Furthermore,~\cite{WulCavRee:2018:mor} exploited classical homogenisation theory and cluster-wise constant strains for efficient MOR. 
}

In this work, we propose a simple and efficient hyperreduced strain space MOR approach for RVEs, called Empirical Material Sampling and Linearisation (EMSL). The approach is conceptually similar to E3C-like {Galerkin} schemes, but also draws on ideas from previous work on incremental structure-preserving strain-space techniques to achieve rapid evaluations in the online phase. As in E3C~\cite{WulHau:2025:ech}, we group the material domain into clusters of similar (strain-space) behaviour, which need not be spatially contiguous. Contrary to previous work~\cite{RuaJuChe:2024:CPR,WulHau:2025:ech,wul:2025:ECC,WulHau:2025:EHR}, we estimate the expected average strain in each of these clusters once per load increment. 
Modes obtained via the Proper Orthogonal Decomposition (POD) allow us to compute a {linearised estimate} of the stress response in the remainder of the cluster based on the stress and material tangent computed at these reference strains\footnote{Michel and Suquet~\cite{MicSuq:2016:mam} employed a similar quadratic approximation of the elastic potential in the context of NTFA.}. This linearisation means that the reduced problem becomes affine in each load step, allowing integration to be performed exactly via operations which can be preprocessed in the offline phase~\cite{BarMadNguPat:2004:eim}. Furthermore, the linearity of the resulting reduced equation systems {means that no Newton iterations are required in the online phase}. 
Since we do therefore not approximate integrals via sample points, but approximate the integrand, our strategy is also related to Empirical Interpolation Methods (EIM)~\cite{BarMadNguPat:2004:eim}, with the caveat that we estimate the nonlinear stress response via a Taylor expansion around an empirically estimated expected reference strain.
{Overall, EMSL can therefore be viewed as exactly integrating an approximation of the material behaviour, while E3C and related schemes approximately integrate the exact material law. Alternatively, we could succinctly describe EMSL as Galerkin projection applied to the load-step-wise affine problem obtained by a cluster-wise linearisation of the material behaviour around an empirically estimated reference point.}

In this work, we show that the new EMSL strategy results in excellent performance in terms of robustness, validation accuracy, and particularly online speed for an example large-deformation hyperelastic RVE problem with {a complex microstructure consisting of interconnected pores}.

We begin this work with a brief overview of computational homogenisation for hyperelastic RVEs in Section~\ref{s:homog}. Then, we discuss strain-space approaches to model order reduction using the POD-Galerkin framework in Section~\ref{s:HR}. In Subsection~\ref{s:ECSW}, we cast a hybrid of the Empirical Cubature Method (ECM)~\cite{HerCaiFer:2017:dhn} and Energy Conserving Weighting and Sampling (ECSW)~\cite{FarAveCha:2014:drn,AnKimDou:2008:OCE} in strain space.
Then, we recall Wulfinghoff et al.'s E3C~\cite{WulHau:2025:ech} in Subsection~\ref{s:E3C}. 
Finally, we describe the EMSL approach in Section~\ref{s:EMSL}. 
{The performance of all methods on an example porous hyperelastic microstructure is investigated in Section~\ref{s:application}.}

\section{Computational homogenisation in strain space}\label{s:homog}

Computational homogenisation approaches such as the multiscale Finite Element (FE\textsuperscript{2}) method postulate an RVE to model the microstructure of a macroscopic component and its mechanical behaviour. The evaluation of a material law on the macroscale can then be replaced by the solution of an RVE problem followed by the homogenisation of stress and stiffness~\cite{Sch:2014:nth}. %Since a fully coupled FE\textsuperscript{2} simulation thus necessitates hundreds of RVE simulations, computational costs must be reduced dramatically, e.g. via MOR techniques. 
In this Section, we give a brief outline of these solution and homogenisation steps. %in the context of hyperelasticity. 
Macroscale quantities are denoted by an overbar $\bar{\cdot}$. More details can be found in the vast reference literature, e.g. in~\cite{Sch:2014:nth,Sch:2000:hnk,SaeSteJav:2016:ach}.

On the microscale, the weak form of the balance of linear momentum, neglecting bulk loads, can be written as 
\begin{equation}
    \int_{\Omega} \delta \vec{F}(\vec{X}) :\vec{P}(\vec{X}) \text{d}V   = 0\,,\label{eq:weakform}
\end{equation}
in terms of the deformation gradient $\vec{F}$ and the first-Piola Kirchhoff stress $\vec{P}$ at some point $\vec{X}\in \Omega$ in the RVE~$\Omega$~\cite[p.82]{Wri:2001:nf}. Hyperelastic material models employ a material law $\vec{P}= \vec{P}(\vec{F})$ which can be derived from a stored energy $W$ as $\vec{P} = \frac{\partial W}{\partial \vec{F}}$, with the nominal stiffness $\ten{\mathbb{A}}=\frac{\partial \vec{P}}{\partial \vec{F}}$~\cite[p.40]{Wri:2001:nf},~\cite[p.207]{Hol:2002:nsm}. 

In classical computational homogenisation, which assumes scales to be clearly separated, the behaviour of the micro- and macroscale are coupled via the Hill-Mandel condition. This condition states that the work done by stress $\vec{\bar{P}}$ and {deformation gradient rate $\vec{\dot{\bar{F}}}$} on a point on the macroscale ought to equal the work done on the RVE via the corresponding {microscale stresses $\vec{{P}}$ and deformation gradient rate $\Delta \vec{\dot{{F}}}$~\cite{Sch:2014:nth}} 
{
\begin{equation}
    \vec{\Bar{P}} : \vec{\dot{\Bar{F}}} = \frac{1}{V} \int_{ \Omega} \vec{{P}} : \vec{\dot{F}} \text{d}V\,.\label{eq:hillmandel}%  \frac{1}{V} \int_{\partial \Omega} \vec{t} \cdot \Delta \vec{x} dA
\end{equation}
}
The macroscopic (i.e., average) deformation gradient $\vec{\bar{F}}$ and stress $\vec{\bar{P}}$ appearing in the above are defined via the integral coupling relations
\begin{equation}
    \vec{\Bar{F}} =\frac{1}{V} \int_{\partial \Omega} \vec{x} \otimes \vec{N} \text{d}A \,,\quad \text{and} \quad  \vec{\Bar{P}} = \frac{1}{V} \int_{\Omega} \vec{P} \text{d}V\,,  \label{eq:FPmac}
\end{equation}
with RVE volume $V$, {outer normal $\vec{N}$ and microscale position $\vec{x}$ in the current configuration}, assuming no inner pore pressure to act within the RVE~\cite{Sch:2000:hnk}.

As is typical in mechanical multiscale frameworks, we model microstructural effects via a periodic RVE, which can be subjected to periodic boundary conditions in the deformation gradient fluctuation 
\begin{equation}
    \vec{\tilde{F}} = \vec{F}-\vec{\bar{F}}\,,
\end{equation}
to satisfy Eq.\eqref{eq:hillmandel} in a physically sensible way. As a consequence, the boundary tractions become anti-periodic~\cite{Sch:2014:nth}.

The mechanical behaviour of the RVE is now parameterised in terms of the macroscopic deformation gradient $\vec{\bar{F}}$, such that the weak form can %more naturally 
be written as 
\begin{equation}
    \int_{\Omega} \delta \vec{\tilde{F}}(\vec{X}) : \vec{P}(\vec{X})  \text{d}V   = 0\,,\quad \text{subject to periodic BCs with prescribed } \vec{\Bar{F}}\,.\label{eq:weakformfluc}
\end{equation}
Eq.~\eqref{eq:weakformfluc} could be discretised with an FE Ansatz and solved using standard solvers, or tackled using FFT-based schemes. The former usually treat the displacement fluctuation $\vec{\tilde{u}}$ as the primary unknown variable, such that the strain fluctuation
\begin{equation}
    \vec{\tilde{F}} = \frac{\partial \vec{\tilde{u}}}{\partial \vec{X}} + \ten{I} 
\end{equation}
fulfills compatibility by construction~\cite{Sch:2014:nth}. Since we only use strain-space solvers in our reduced models, and the snapshot data used to train these spans a linear subspace in which compatibility is fulfilled, there is no need to construct an additional enforcement mechanism here.

Once the RVE problem has been solved, the macroscopic first Piola-Kirchhoff stress $\vec{\Bar{P}}$ can be postprocessed from the stress field on the RVE by volume averaging (see Eq.~\eqref{eq:FPmac}). The macroscopic nominal stiffness $\ten{{\bar{\mathbb{A}}}}$ is then defined as the rate of the change of $\vec{\Bar{P}}$ with $\vec{\Bar{F}}$, i.e. $\ten{{\bar{\mathbb{A}}}} = \frac{\partial \vec{\Bar{P}}}{\partial \vec{\Bar{F}}}$~\cite{Sch:2014:nth}.

In practice, it is advantageous to pose the RVE problem not in terms of the deformation gradient $\vec{\bar{F}}\in \mathbb{R}^{3\times 3}$, but in terms of the symmetric right stretch tensor $\vec{\bar{U}}\in \mathbb{R}_{\text{sym}}^{3\times 3}$. This reduces the parameter space in which to sample for snapshot data, necessitates only six rather than %nice
nine perturbations for tangent computation, and ensures rotation invariance of reduced models~\cite{KocBreWul:2018:CEO,LanHutKie:2024:mhr,Wul:2025:HHU}. 
As a consequence, the stress measure to be homogenised becomes the Biot stress. For the sake of simplicity, and since the rotations encountered in the example considered here are not excessively large, the RVE problem is still posed in terms of the deformation gradient in this work. In any case, the MOR techniques discussed below can readily be implemented in homogenisation codes working with the Biot stress as well.

\section{Strain space hyperreduction}\label{s:HR}

\subsection{POD-Galerkin in strain space}

\begin{figure}
    \centering
    \includegraphics[width=0.6\linewidth,trim={2cm 8cm 2cm 6cm},clip]{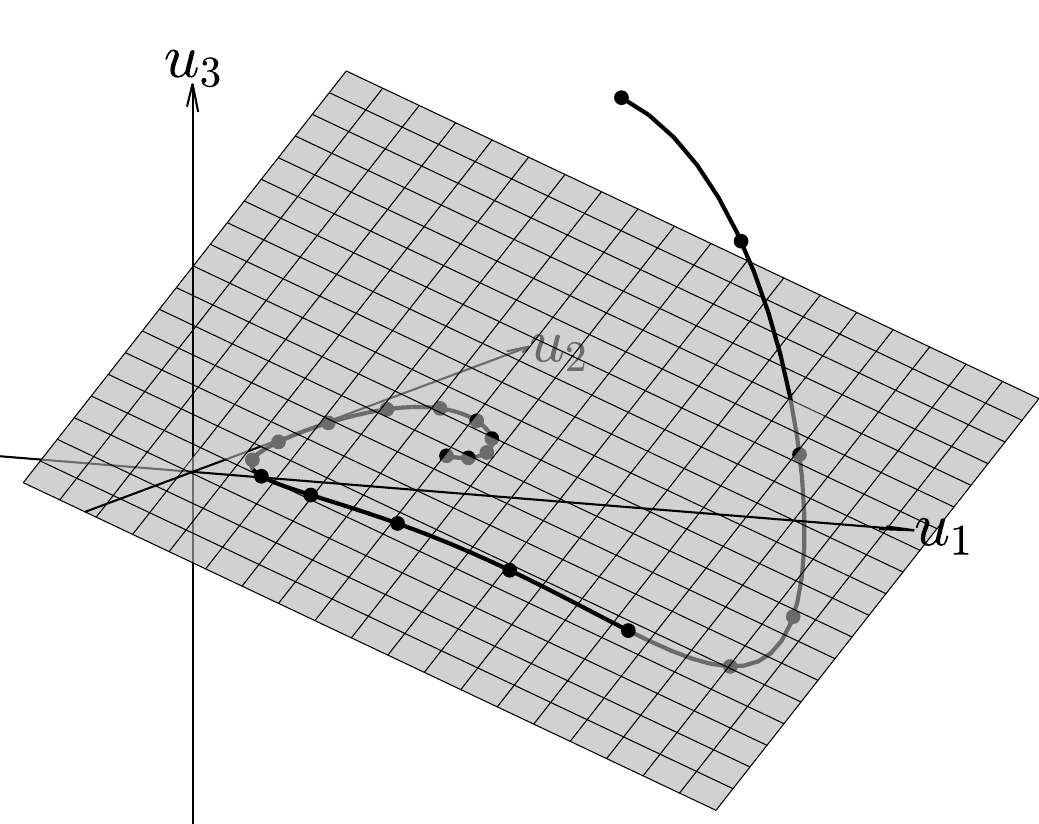}
    \caption{Conceptual illustration of snapshots (black dots) on a $\delta=1$-dimensional solution manifold (black line) and a $d=2$-dimensional linear POD subspace (gray plane) in a solution space originally discretised via $D=3$ unknowns $u_i$.}
    \label{fig:POD}
\end{figure}

Classical FE- or FFT-based homogenisation techniques discretise the primary unknown field on the RVE -- which might be the displacement $\vec{\Tilde{u}}(\vec{X})$ or {deformation gradient} $\vec{\Tilde{{F}}}(\vec{X})$ fluctuation -- using a large number (e.g. $D\approx100,000$ {for a real microstructure}) of unknown nodal values. These can flexibly represent solutions on a complex microstructure, but the resulting solution schemes incur high computational costs, because large nonlinear equation systems have to be solved iteratively for these $D$ values. Additionally, the accurate integration of the weak form for varied solution fields requires a large number of integration points $G\propto D$ at which the material law has to be evaluated. {Even when statistically similar RVEs are used, the computational costs (with e.g. $D\approx10,000$ for a dual-phase steel microstructure) remain significant~\cite{SchBalBra:2015:d3s}.}
However, the parametric setting allows the vast space of possible solutions to be reduced significantly. The set of %all
possible solutions to a hyperelastic RVE problem for arbitrary macroscopic deformation gradients $\vec{\bar{{F}}}\in \mathbb{R}^{3\times 3}$ is a $\delta=6$-dimensional solution manifold $\mathcal{M}_{\vec{\Tilde{{F}}}}=\{\vec{\Tilde{{F}}}\mid \exists \vec{\bar{{F}}}: \int_{\Omega} \delta \vec{\Tilde{{F}}}(\vec{X}) : \vec{{P}}(\vec{X}) \text{d}V   = 0\}$~\cite{DanCasAkk:2020:mor,DanCasAkk:2022:pca,SchFau:2025:MLA}. The solution manifold is 6- and not 9-dimensional on account of the aforementioned rotation invariance of the deformation gradient fluctuation; see Section~\ref{s:homog}. {Fig.~\ref{fig:POD} conceptually illustrates a $\delta=1$-dimensional solution manifold in a $D=3$-dimensional solution space.}

The popular POD-Galerkin framework exploits this opportunity for dimension reduction. It postulates an Ansatz which approximates the unknown field -- in this case, the deformation gradient fluctuation $\vec{\Tilde{\underline{F}}}$ in Voigt notation 
\begin{equation}
    \vec{\Tilde{\underline{F}}}(\vec{X}) = \sum_{i=1}^d y_i \ten{\varphi}_i(\vec{X})\label{eq:GalerkinAnsatz}
\end{equation}
using global mode functions $\ten{\varphi}_i(\vec{X})$ and scalar coefficients $y_i$. This approximation spans a linear, $d>\delta =6$-dimensional linear subspace of the solution space, which in the ideal case is a tight superspace of the $6$-dimensional solution manifold. The solution field can thus be discretised in terms of $d\ll D$ rather than $D$ unknowns, which accelerates the solution of equation systems significantly. A conceptual illustration of a $d=2$-dimensional POD approximation of a $\delta = 1$-dimensional solution manifold in a $D=3$-dimensional solution space can be found in Fig.~\ref{fig:POD}. As can be intuited from this sketch, the approximation quality of the POD depends on the worst-case normal-direction distance from the approximation space (the gray plane in the figure) to the solution manifold (the black line), which can be formalised as the Kolmogorov n-width~\cite{Peh:2022:bkb}.

We discuss how to practically compute the modes $\vec{\varphi}_i(\vec{X})$ from preliminary simulations (for example, using the FEM) below.  In order to deal with vector- rather than tensor-valued unknown quantities, we use Voigt notation throughout the remainder of this text; all tensors written in Voigt notation will be denoted with an underline $\underline{\cdot}$.
%An appropriate choice of $\varphi_i(\vec{X})$ in the discrete setting is discussed below. 

We can substitute the Galerkin Ansatz into the weak form in Eq.~\eqref{eq:weakformfluc} to obtain
%When substituting such a Galerkin Ansatz into the weak form in Eq.~\eqref{eq:weakformfluc}, we obtain
\begin{equation}
    \sum_{i=1}^d \delta y_i \int_{\Omega} \vec{\varphi}_i^{T}(\vec{X}) \vec{\underline{P}}(\vec{\underline{F}}(\vec{X})) \text{d}V   = 0\,.\label{eq:Galerkin}
\end{equation}
When evaluated at appropriately chosen numerical integration points {$g\in G$}, Eq.~\eqref{eq:Galerkin} becomes approximately
\begin{equation}
    \sum_{i=1}^d \delta y_i \sum_{g\in G} \vec{\varphi}_i^{T}(\vec{X}^g) \vec{\underline{P}}(\vec{\underline{F}}(\vec{X}^g)) V^g = \sum_{i=1}^d \delta y_i \sum_{g\in G} \vec{\varphi}_i^{gT} \vec{\underline{P}}^{g} V^g = 0\,,\label{eq:FGalerkin}
\end{equation}
where $V^g$ denotes the volume associated with integration point $g$ at coordinate $\vec{X}^g$, and the superscript $g$ indicates discrete values at these integration points. For simplicity of presentation, we collect the mode coefficients $y_i$ in a vector $\vec{y}$ and define associated discrete mode matrices $\ten{\psi}^g\in \mathbb{R}^{9\times d}$,
\begin{equation}
    \ten{\psi}^g = [\ten{\varphi}_1(\vec{X}^g),...,\ten{\varphi}_d(\vec{X}^g)]\,,
\end{equation}
such that discretised weak form can be written as
\begin{equation}
    \partial \vec{y}^T \sum_{g\in G} \ten{\psi}^{gT} \vec{\underline{P}}^{g} V^g = 0\,.\label{eq:weakGalerkin}
\end{equation}

In the POD-Galerkin framework, the discrete mode matrices
\begin{equation}
    \ten{\psi} =
    \begin{bmatrix}
        \ten{\psi}^1\\
        \vdots \\
        \ten{\psi}^{|G|} 
    \end{bmatrix}
    \in \mathbb{R}^{9|G|\times d}
\end{equation} 
can be identified from $s$ strain snapshots $\vec{\underline{F}}^{gj}, j \in \{1,...,s\}$ of the unknown field at the integration points $g\in G$. These can be collected in a matrix
\begin{equation}
    \ten{\underline{F}}^s = 
    \begin{bmatrix}
        \vec{\underline{F}}^{11} & \vec{\underline{F}}^{12} & \dots & \vec{\underline{F}}^{1s} \\
        \vec{\underline{F}}^{21} & \vec{\underline{F}}^{22} & \dots & \vec{\underline{F}}^{2s} \\
        \vdots &  \vdots & \ddots & \vdots \\
        \vec{\underline{F}}^{|G|1} & \vec{\underline{F}}^{|G|2} & \dots & \vec{\underline{F}}^{|G|s}
    \end{bmatrix}
\end{equation}
and subjected to a singular value decomposition
\begin{equation}
    \ten{\underline{F}}^s = \ten{L} \ten{\Sigma} \ten{R}\,.
\end{equation}
Then, the $d$ the leading left singular vectors define the discrete global mode matrix~\cite{Pea:1901:llp}
\begin{equation}
    \ten{\psi}=[\vec{L}^j]_{j\in[1,..d]}\,.
\end{equation}
Solving the weak form in Eq.~\eqref{eq:FGalerkin} with this basis recovers the strain-space reduced basis method by Kunc et al.~\cite{KunFri:2019:fsh}.

\subsection{Empirical Cubature (ECM) and Energy Conserving Weighting and Sampling (ECSW)}\label{s:ECSW}

The POD-Galerkin Ansatz reduces the number of unknown variables of the RVE problem to $d\ll D$, which considerably accelerates the solution of linear systems required by the iterative nonlinear equation solver. However, integrating the weak form still requires $|G|$ evaluations of the material law in each iteration, which is a crucial contributor to computational cost especially in moderately-sized problems with $D <100,000$.
Popular hyperreduction approaches such as Energy Conserving Sampling and Weighting (ECSW) or the Empirical Cubature Method (ECM) approximate the integral appearing in the Galerkin-reduced weak form in Eq.~\eqref{eq:weakGalerkin} by summing over a reduced subset $H \subset G$ of the integration points of the original high-fidelity model~\cite{AnKimDou:2008:OCE,FarAveCha:2014:drn,HerCaiFer:2017:dhn,Her:2020:MMP}.

Here, we deploy a strain-space variant of ECM which also draws on some ideas taken from ECSW\footnote{Alternatively, we could describe this approach as ECSW with ECM-like modifications. ECSW and ECM are closely related, with the key differences being that ECM operates on the Gauss point level and centers its snapshot data. Different solvers are used in the training phase and ECM introduces an additional volume constraint.}, with the integration rule written, on the Gauss point level, as
\begin{equation}
    \partial \vec{y}^T \sum_{c\in H} \ten{\psi}^{cT} \vec{\underline{P}}^{c}(\vec{\underline{F}}^{c}) \xi^c = 0\,,\label{eq:ecsw}
\end{equation}
where $|H|\ll |G|$ and the $\xi^c$ are non-negative coefficients which can be interpreted as effective volumes associated with the selected integration points. {The volumes $\xi^c$ are thus enlarged relative to the original $V^g$ to compensate for neglected integration points. }

When cast in strain space, the cubature scheme requires the collection of $s$ integration point stress snapshots $\ten{\underline{P}}^{gj}, g \in G, j \in 1,..,s$ and original integration point volumes $V^g$. Then, the reduced set of integration points and associated volumes can be computed via the sparse nonnegative least squares (NNLS) problem
\begin{equation}
    \min_{\xi^c,H} \sum_{j=1}^s \| ( \sum_{g\in G} \ten{\psi}^{gT}\vec{\underline{P}}^{gj} V^g-\sum_{c\in H} \ten{\psi}^{cT} \vec{\underline{P}}^{cj} \xi^c) \|^2 \quad \text{s.t.} \quad H\subset G, \quad |H|\overset{!}{=}m\,. \label{eq:NNLS_ECSW}
\end{equation}
As in ECSW, Eq.~\eqref{eq:NNLS_ECSW} can physically be interpreted as an energy criterion: the virtual work done by the strain modes on the sampled and weighted integration point stresses should come as close as possible to the work done by the same modes on the original Gauss point stresses.
This can be written in matrix form as a sparse linear least squares problem with nonegativity constraints. 
Eq.~\eqref{eq:NNLS_ECSW} can be solved to a specified tolerance or until $m=|H|$ nonzero integration points and coefficients $\xi^c$ have been identified. Here, we identify {the reduced set of} integration points using the Lawson-Hanson solver also applied in ECSW~\cite{FarAveCha:2014:drn}. The computational cost incurred in this offline procedure is moderate. For further details, we refer the reader to~\cite{AnKimDou:2008:OCE,FarAveCha:2014:drn,HerCaiFer:2017:dhn,Her:2020:MMP}.

In the spirit of ECM, we also impose a volume constraint on the weights $\xi^c$ using a penalty term with penalty parameter $p_\text{vol}$, such that the NNLS problem becomes
\begin{align}
    \min_{\xi^c,H} & \sum_{j=1}^s \| ( \sum_{g\in G} \ten{\psi}^{gT}\vec{\underline{P}}^{gj} V^g-\sum_{c\in H} \ten{\psi}^{cT} \vec{\underline{P}}^{cj} \xi^c) \|^2 + p_\text{vol}(\sum_{c\in H} \xi^c - \sum_{g \in G} V^g)^2 \quad \nonumber \\ & \text{s.t.} \quad H\subset G, \quad |H|\overset{!}{=}m\,. \label{eq:volconst}
\end{align}
As observed in~\cite{HerCaiFer:2017:dhn}, the first term in Eq.~\eqref{eq:NNLS_ECSW}, which measures the virtual work done by the modes on Gauss point strains, vanishes to the solver tolerance when converged stresses are used for training. In this case, the volume constraint also helps avoid degenerate solutions.

For accurate homogenisation, additional (independent) conditions encouraging accurate stress integration can be introduced into the NNLS problem, finally yielding
\begin{align}
    \min_{\xi^c,H} & \sum_{j=1}^s \| ( \sum_{g\in G} \ten{\psi}^{gT}\vec{\underline{P}}^{gj} V^g-\sum_{c\in H} \ten{\psi}^{cT} \vec{\underline{P}}^{cj} \xi^c) \|^2 + p_\text{vol}(\sum_{c\in H} \xi^c - \sum_{g \in G} V^g)^2\nonumber\\
    &  + \sum_{j=1}^s \| ( \vec{\bar{\underline{P}}}^{j}-\sum_{c\in H} \xi^c \vec{\underline{P}}^{cj}) \|^2\quad \text{s.t.} \quad H\subset G, \quad |H|=m \,.\label{eq:NNLShomog}
\end{align}
Alternatively, separate integration weights can be computed for the homogenisation procedure. This may improve the accuracy of the stress computation, but slightly violates the consistency of the algorithmic tangent, which is derived below.

Once the NNLS problem has been solved and the reduced integration points identified, Eq.~\eqref{eq:ecsw} can be solved via a Newton-Raphson scheme. The residual
\begin{equation}
    \vec{r}(\vec{y};\vec{\bar{\underline{F}}}) = \sum_{c\in H} \ten{\psi}^{cT} \vec{\underline{P}}(\vec{\underline{F}}^{c}) \xi^c\,, \quad \text{where} \quad \vec{\underline{F}}^{c} = \vec{\bar{\underline{F}}} + \vec{\underline{\tilde{F}}}^{c}\,,\quad \vec{\underline{\tilde{F}}}^{c} = \ten{\psi}^c \vec{y}\label{eq:ECMres}
\end{equation}
can be differentiated with respect to the reduced unknown variables to yield the tangent
\begin{equation}
    \ten{K}(\vec{y};\vec{\bar{\underline{F}}}) = \frac{\partial \vec{r}}{\partial \vec{y}} = \sum_{c\in H} \ten{\psi}^{cT} \frac{\partial \vec{\underline{P}}}{\partial \vec{\underline{F}}^{c}} \frac{\partial \vec{\underline{F}}^{c}}{\partial \vec{y}} \xi^c = \sum_{c\in H} \ten{\psi}^{cT} \ten{\mathbb{\underline{A}}}^{c} \ten{\psi}^c \xi^c\,,
\end{equation}
and solutions can be sought via
\begin{equation}
    \ten{K}(\vec{y};\vec{\bar{\underline{F}}}) \Delta \vec{y} = - \vec{r}(\vec{y};\vec{\bar{\underline{F}}})\,,
\end{equation}
and $\vec{y} \leftarrow \vec{y}+\Delta \vec{y}$. 
% As a consequence of the POD Ansatz, the value of the strain $\vec{\underline{F}}^c$ at the integration points can be computed from the reduced variables using Eq.~\eqref{eq:GalerkinAnsatz},
% \begin{equation}
%     \vec{\underline{F}}^{c} = \vec{\bar{\underline{F}}} + \ten{\psi}^c \vec{y}\,.
% \end{equation}

Once an approximate solution has been found, the stress can be homogenised using the reduced integration weights obtained {using Eq.~\eqref{eq:NNLShomog}}
\begin{equation}
    \vec{\bar{\underline{P}}} = \frac{1}{V} \sum_{c\in H} \xi^c \vec{\underline{P}}^{c}\,.
\end{equation}
Consequently, the associated algorithmic tangent is given by
\begin{equation}
    \ten{\bar{\mathbb{\underline{A}}}} = \frac{\partial \vec{\bar{\underline{P}}}}{\partial \vec{\bar{\underline{F}}}} = \frac{1}{V} \sum_{c\in H} \xi^c \frac{\partial \vec{\underline{P}}^{c}}{\partial \vec{\underline{F}}^{c}} \frac{\partial \vec{\underline{F}}^{c}}{\partial \vec{\bar{\underline{F}}}}= \frac{1}{V} \sum_{c\in H} \xi^c \ten{\mathbb{\underline{A}}}^{c} \frac{\partial (\vec{\bar{\underline{F}}} + \vec{\tilde{\underline{F}}}^{c})}{\partial \vec{\bar{\underline{F}}}}\,,
\end{equation}
and thus
\begin{equation}
    \ten{\bar{\mathbb{\underline{A}}}}^ v = \frac{1}{V} \sum_{c\in H} \xi^c \ten{\mathbb{\underline{A}}}^{c} + \frac{1}{V} \sum_{c\in H} \xi^c \ten{\mathbb{\underline{A}}}^{c} \frac{\partial \vec{\tilde{\underline{F}}}^{c}}{\partial \vec{\bar{\underline{F}}}} = 
    \frac{1}{V} \sum_{c\in H} \xi^c \ten{\mathbb{\underline{A}}}^{c} + \frac{1}{V} \sum_{c\in H} \xi^c \ten{\mathbb{\underline{A}}}^{c} \ten{\psi}^c \frac{\partial \vec{y}}{\partial \vec{\bar{\underline{F}}}}\,.
\end{equation}

The sensitivity $\frac{\partial \vec{y}}{\partial \vec{\bar{\underline{F}}}}$ can be derived using the classical argument which perturbs the discretised weak form~\cite{Sch:2014:nth}
\begin{equation}
    \sum_{c\in H} \partial \vec{\tilde{\underline{F}}}^{cT} \ten{\mathbb{\underline{A}}}^c \Delta \vec{\underline{F}}^c \xi^c=0 \,,
\end{equation}
which implies
\begin{equation}
    \delta \vec{y}^T \sum_{c\in H} \ten{\psi}^{cT} \ten{\mathbb{\underline{A}}}^{c} (\Delta \vec{\tilde{\underline{F}}}^{c} + \Delta \vec{\bar{\underline{F}}}) \xi^c = \delta \vec{y}^T (\sum_{c\in H} \ten{\psi}^{cT} \ten{\mathbb{\underline{A}}}^{c} \ten{\psi}^c \xi^c \Delta \vec{y}  + \sum_{c\in H} \ten{\psi}^{cT} \ten{\mathbb{\underline{A}}}^{c} \xi^c \Delta \vec{\bar{\underline{F}}}) = 0\,,
\end{equation}
i.e.
\begin{equation}
    \ten{K}\Delta \vec{y} = - \ten{L} \Delta \vec{\bar{\underline{F}}}\,,
\end{equation}
with 
\begin{equation}
    \ten{L} = \sum_{c\in H} \ten{\psi}^{cT} \ten{\mathbb{\underline{A}}}^{c} \xi^c\,,
\end{equation}
such that
\begin{equation}
    \frac{\partial \vec{y}}{\partial \vec{\bar{\underline{F}}}} = - \ten{K}^{-1} \ten{L}\,.
\end{equation}

Consequently, only $|H|\ll|G|$ of the original Gauss points need to be queried in every Newton-Raphson step in the online phase, which reduces assembly times considerably. A minimal pseudocode for the online phase of strain space ECM can be found in Algorithm~\ref{alg:ECSW_online}. As the nonlinear equation systems to be assembled and solved are low-dimensional (on the order of tens or hundreds of degrees of freedom $d$), the dominant contributors to computational cost in the reduced model tend to be the number of evaluations of the material routine at the reduced integration points (of which there are $m$, also in the order of tens or hundreds). Further runtime optimisation must therefore address this bottleneck directly or indirectly, while maintaining satisfactory levels of accuracy.

\begin{algorithm}[h!]
\caption{Strain space ECM or E3C online phase}

\For{$\ten{\bar{\underline{F}}}$ in loadpath}{
\While{$\|\vec{r}\|<tol$}{

\tcp{Solve for increment}
$\Delta \vec{y} \gets \text{solve}(\ten{K}\vec{y}=-\vec{r} )$\;
\tcp{Increment}
${\Vec{y}} \gets {\Vec{y}} + \Delta {\Vec{y}}$\; 
${\Vec{\underline{F}}}^{c} \gets {\Vec{\underline{F}}}^{c} + \ten{\psi}^c \Delta {\Vec{y}}$\; 

\tcp{Integration point stress and stiffness}
$\vec{\underline{P}}^{c}, \ten{\mathbb{\underline{A}}}^{c} \gets \text{mat}(\ten{\underline{F}}^{c}), c \in H$\;
\tcp{Assembly}
$\vec{r} \gets \sum_{c\in H} \ten{{\psi}}^{cT} \vec{{\underline{P}}}^{c} \xi^c$\;
$\ten{K} \gets \sum_{c\in H} \ten{{\psi}}^{cT} \ten{{\mathbb{\underline{A}}}}^{c} \ten{{\psi}}^{c} \xi^c$\;

}

\tcp{Homogenise stress}
$\ten{\bar{\underline{P}}} \gets \sum_{c\in H}  \vec{{\underline{P}}}^{c} \xi^c$\;

\tcp{Algorithmic consistent tangent}
$\ten{L} \gets \sum_{c\in H} \ten{\psi}^{cT} \ten{\mathbb{\underline{A}}}^{c} \xi^c$\;
$\ten{X} \gets \text{solve}(\ten{K}\ten{X} = -\ten{L})$\;
$\ten{\bar{\mathbb{\underline{A}}}} \gets \frac{1}{V} \sum_{c\in H} \ten{\mathbb{\underline{A}}}^{c} \xi^c - \frac{1}{V}\ten{L} \ten{X}$

}
\label{alg:ECSW_online}
\end{algorithm}

\subsection{Empirically Corrected Cluster Cubature (E3C)}\label{s:E3C}

ECM and ECSW pick the integration points to be used in the hyperreduced model among the Gauss points of the original high-fidelity model. As noted by Hernandez et al.~\cite{HerBraPar:2024:CCE} and Wulfinghoff et al.~\cite{WulHau:2025:ech}, however, hyperreduced integration points $H$ need not be a subset $H\subset G$ of these original Gauss points. Indeed, such a choice restricts the search for integration points to a -- possibly quite limited -- discrete subset of the possible search space.

E3C circumvents this constraint by searching for strain bases $\ten{\psi}^c, c \in H$ corresponding to arbitrary reduced integration points $H$, which result in accurate integration of the residual and homogenisation~\cite{WulHau:2025:ech}. These are initialised using a clustering procedure, which identifies possible mode samples to characterise the strain distribution. While clustering yields integration points which can closely match the statistical attributes of the strain distribution, the resulting integration rule is not physically consistent. In a second optimisation step, E3C therefore enforces consistency by ensuring the virtual work vanishes at snapshots and the homogenised stress is captured accurately.

For initialisation, E3C applies a k-means algorithm to the Gauss point bases $\ten{\psi}^g, g \in G$ to obtain clusters $C\subset G$ of Gauss points, where $C$ denotes the set of all integration points associated with a cluster. Centroids $\ten{\psi}^c, c \in H$ of these clusters can be computed as
\begin{equation}
    \ten{\psi}^c = \frac{1}{\xi^c} \sum_{g\in C} \ten{\psi}^g V^g\,,
\end{equation}
where $\xi^c$ is the cluster volume 
\begin{equation}
    \xi^c = \sum_{g\in C} V^g\,.
\end{equation}

Then, the $\ten{\psi}^c, c \in H$ initialised via clustering are treated as design variables to minimise the snapshot residuals as well as the deviation from the homogenised stress snapshots
\begin{align}
    \min_{\ten{\psi}^c} & \sum_{j=1}^s \|\sum_{c\in H} \ten{\psi}^{cT} \vec{\underline{P}}(\vec{\underline{F}}^{cj}) \xi^c \|^2 + \sum_{j=1}^s \|\sum_{c\in H} \vec{\underline{P}}(\vec{\underline{F}}^{cj}) \xi^c - \vec{\bar{\underline{P}}}^{j} \|^2 \quad \nonumber\\
    & + p_{\text{strain}} (\sum_{c\in H} \vec{\tilde{\underline{F}}}^{c} \xi^c - \sum_{g\in G} \vec{\tilde{\underline{F}}}^{g} V^g)^2\; \quad \text{where} \quad \vec{{\underline{F}}}^{cj} = \vec{\bar{\underline{F}}}^{c} + \ten{\psi}^c \vec{y}^j\,,
\end{align}
subject to a mean fluctuation constraint imposed using a penalty term with coefficient $p_{\text{strain}}$.
E3C thus optimises the integration point modes obtained via clustering for improved fulfillment of the RVE's mechanics in the snapshot set. If we do not perform such an empirical correction, we obtain MOR scheme similar to that proposed in~\cite{RuaJuChe:2024:CPR}, {but we have observed that empirical correction is necessary for competitive performance on examples with complicated microstructures.} %but in the example explored below, empirical correction is necessary for competitive performance. 
{Our implementation of E3C's training phase differs slightly from the original method proposed in~\cite{WulHau:2025:ech}: rather than a Conjugate Gradient scheme, we employ L-BFGS for minimisation. Furthermore, Wulfinghoff et al.~\cite{WulHau:2025:ech} deploy an optimised data collection scheme which samples for additional snapshots in regions with comparatively higher E3C errors. For the sake of comparability, we instead train all models with the same, fixed set of snapshots. }%E3C performs well none the less
Note that E3C fulfills the volume constraint in Eq.~\eqref{eq:volconst} exactly by construction, as integration point weights are computed as sums over the Gauss point weights associated with each cluster.

The fitting of bases to the snapshot residuals and homogenised stresses is comparatively expensive. However, all operations following the strain basis computation and clustering are performed in the reduced space, mitigating memory costs. Contrary to ECSW or ECM, stress  or residual snapshots on the microscale are not required. Furthermore, the physics-informed optimisation of strain-space integration points allows E3C to sample a much larger search space and to find improved integration points in comparison with ECSW or ECM schemes, which search only discrete points in strain space. {We refer the interested reader to} Hernandez et al.~\cite{HerBraPar:2024:CCE}, where a similar observation is made. Therefore, E3C can obtain reduced models with fewer integration points $m$ while maintaining the accuracy achieved by ECSW or ECM.

The reduced E3C residual, stiffness matrix, homogenised stress, and algorithmically consistent tangent can be written completely analogously to those of strain space ECM in Subsection~\ref{s:ECSW}, with the difference being that the integration points are no longer a subset of the integration points of the original high-fidelity model. We therefore do not repeat these equations here. The pseudocode in Algorithm~\ref{alg:ECSW_online} is also valid for the online phase of E3C.

\begin{figure}
    \centering
    \def\svgwidth{\linewidth}
    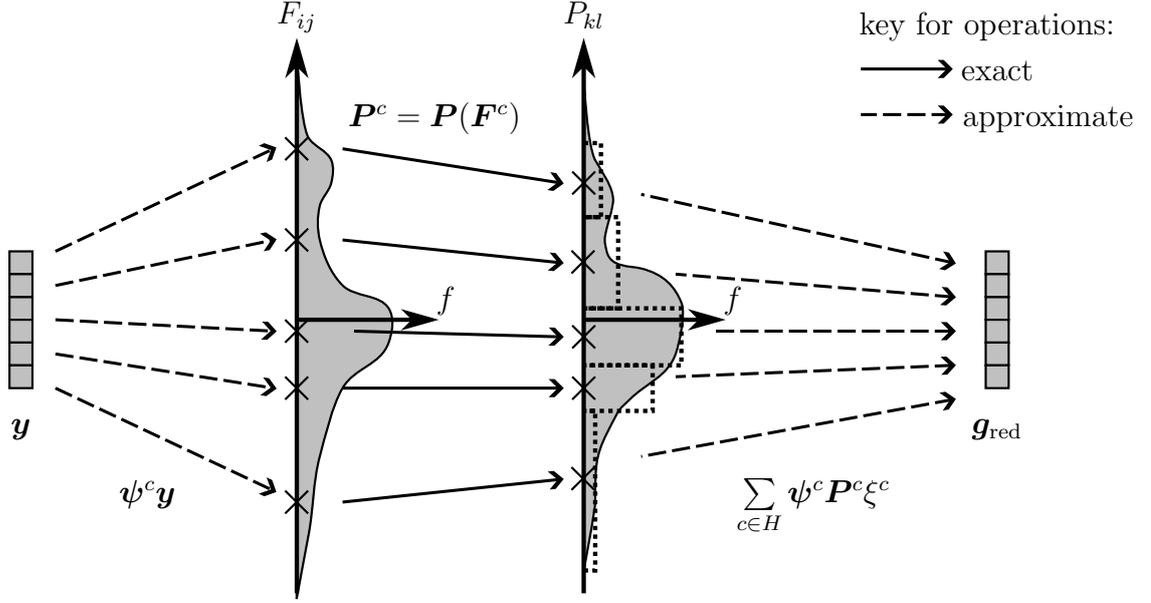
    %\includesvg[width=0.9\textwidth]{reduced_cubature_labels.svg}
    \caption{Illustration of reduced cubature approaches such as ECSW, ECM, or E3C. Strain samples are approximated via integration point mappings from reduced space to strain space. The material law is used to compute the corresponding stresses. Integration is performed approximately, with the samples constituting integration points.}
    \label{fig:RC}
\end{figure}

{In the online phase, E3C effectively pursues an approach to hyperreduction which firstly obtains samples of the strain distribution via a mapping from the reduced space to strain space. Then, the material law is queried to compute corresponding stress samples. These are projected onto the approximation space using the (empirically corrected) bases at these reduced integration points, and the results are weighted with cluster volumes and accumulated. E3C could therefore be said to approximately integrate an exact material law using empirical samples of the strain distribution. Fig.~\ref{fig:RC} illustrates this approach graphically. ECM follows a similar approach, but chooses these samples less freely. Furthermore, displacement-space ECSW or ECM schemes essentially introduce two intermediate steps, as strain samples are obtained from displacement data, and a residual is computed from the stress prior to projection and integration.} 

\section{Empirical Material Sampling and Linearisation (EMSL)}\label{s:EMSL}

In this work, we introduce an approach which, instead of approximately integrating an exact material law, exactly integrates an approximation of the material behaviour. We call this approach Empirical Material Sampling and Linearisation (EMSL).

\begin{figure}
    \centering
    \def\svgwidth{\linewidth}
    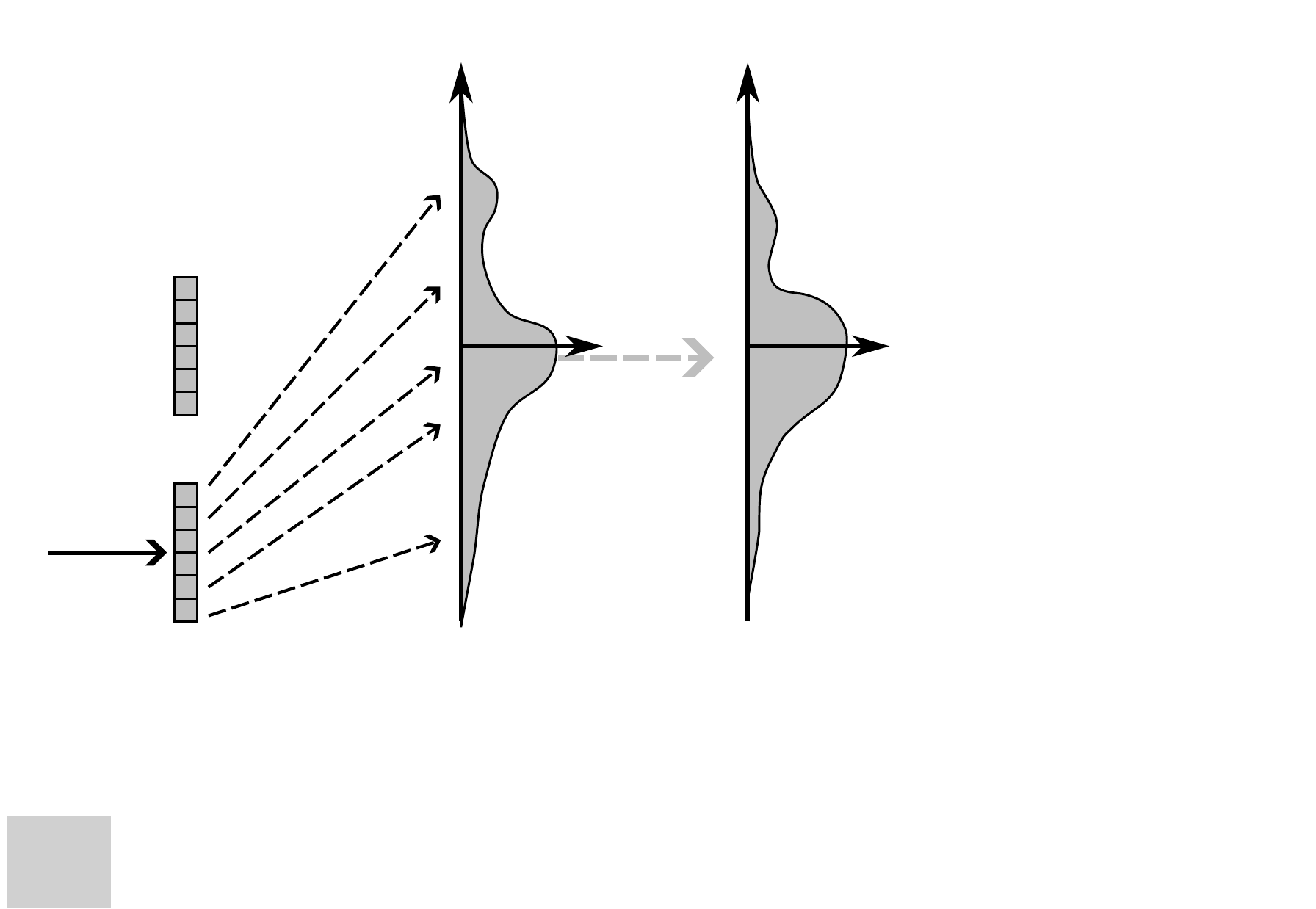
    %\includesvg[width=1.0\textwidth]{emsl_labels_2.svg}
    \caption{Illustration of EMSL{, for a case involving 5 clusters}. There are two mappings from the reduced space to strain space: the estimate $\vec{\bar{y}}$ (obtained from $\vec{\bar{\underline{F}}}$) is used to infer the expected average strain fluctuation {in each of the 5 clusters} around which the material law is linearised. Meanwhile, $\vec{y}$ parameterises the strain fluctuation throughout the {entire domain}. The linearised material law allows for approximate stresses to be computed {throughout each of the 5 clusters}. These approximated stresses can be projected onto the approximation space and integrated exactly, over the entire integration domain.}
    \label{fig:EMSL}
\end{figure}

In contrast to strain space ECM and E3C, EMSL does not aim to select strain-space integration points via which to repeatedly approximate a reduced residual within a Newton-Raphson scheme. Instead, we compute an empirical estimate of cluster-wise reference strains, around which to linearise the material behaviour in each load increment (not each Newton-Raphson step). This results in an affine (i.e. linear) problem in each load step. Alternatively, EMSL can be run iteratively as a fixed-point scheme. 
%Rather than approximately integrating an exact material law, EMSL could therefore be said to exactly integrate an approximate material law. 
This conceptual approach is visualised in Fig.~\ref{fig:EMSL}.
For the reader's convenience, some equations already stated in Sections~\ref{s:ECSW} and~\ref{s:E3C} are repeated in the discussion below.

As in E3C, EMSL firstly groups Gauss points $g\in G$ into $|H|$ clusters {using k-means}, such that each $g\in C\subset G$. The material behaviour in each of these clusters is later linearised around a cluster-wise strain-space reference point, defined as the centroid basis
\begin{equation}
    \ten{\psi}^c = \frac{1}{\xi^c} \sum_{g \in C} \ten{\psi}^g V^g\,.
\end{equation}
$\ten{\psi}^c$ can be interpreted as a basis for the average strain within each cluster. As above, we also have
\begin{equation}
    \xi^c = \sum_{g\in C} V^g\,.
\end{equation}

Contrary to E3C, we seek to estimate the average cluster strain $\vec{\underline{F}}^{c}$ once at the beginning of each load step. The goal of this estimate is to cheaply obtain a reference point around which to linearise the material behaviour within each cluster.
To this end, we fit a simple linear model
\begin{equation}
    \ten{Y}^s \approx \ten{M} \vec{\Bar{\underline{F}}}^{s}\,,
\end{equation}
from parameter space to the reduced space using snapshots $\ten{Y}^s\in \mathbb{R}^{d\times s}$ and $\vec{\Bar{\underline{F}}}^{s}\in \mathbb{R}^{9\times s}$ in both of these spaces. The least-squares optimal $\ten{M}$ can be computed using the pseudoinverse,
\begin{equation}
    \ten{M} = \ten{Y}^s \vec{\Bar{\underline{F}}}^{sT} (\vec{\Bar{\underline{F}}}^{s}\vec{\Bar{\underline{F}}}^{sT})^{-1}\,.
\end{equation}
At the beginning of each load step, we can therefore approximate the average strain in each cluster $\ten{\underline{F}}^c$ by inferring from the macroscopic strain $\vec{\bar{\underline{F}}}$ to a reduced-space state estimate $\vec{\bar{y}}$,
\begin{equation}
    \vec{\bar{y}} = \ten{M} \vec{\bar{\underline{F}}}\,,
\end{equation}
and reconstructing
\begin{equation}
    \vec{\tilde{\underline{F}}}^{c} = \ten{\psi}^c \vec{\bar{y}}\,, \quad \text{such that} \quad  \vec{{\underline{F}}}^{c} =\vec{\bar{\underline{F}}} + \vec{\tilde{\underline{F}}}^{c}\,.\label{eq:pred}
\end{equation}
We use this empirical estimate of the cluster strain average as a reference point around which to linearise the material behaviour in each cluster. {Of course, the estimates $\vec{\bar{y}}$ and $\vec{{\underline{F}}}^{c}$, which result from a simple linear fit, are generally still some distance away from the corresponding high-fidelity ground-truth values. To approach those desired values more closely, EMSL next computes a solution $\vec{y}$ to the reduced equilibrium equations, which are linearised around the reference point given by $\vec{\bar{y}}$ and $\vec{{\underline{F}}}^{c}$. Generally, $\vec{\bar{y}}$ and $\vec{y}$ do not agree.}

To compute the correction $\vec{y}$ to the estimate $\vec{\bar{y}}$, we use a POD-Galerkin model of the strain fluctuation at all integration points {$g \in C \subset G$ in all clusters $C$},
\begin{equation}
    \vec{\tilde{\underline{F}}}^{g} = \ten{\psi}^g \vec{y}\,.
\end{equation}
This is contrary to the approaches outlined in Subsections~\ref{s:ECSW} and~\ref{s:E3C}, which deploy the same Ansatz only to a reduced subset of integration points.
For each integration point $g\in C\subset G$ of the full-order model --  which lies in some cluster $C$ -- we now linearise the material behaviour around the cluster-wise reference strain $\ten{\underline{F}}^{c}$ computed using Eq.~\eqref{eq:pred},
\begin{align}
    \vec{\underline{P}}(\ten{{\underline{F}}}^{g}) &= \vec{\underline{P}}(\ten{{\underline{F}}}^{c})+\ten{\mathbb{\underline{A}}}(\ten{{\underline{F}}}^{c})(\ten{\underline{F}}^{g}-\ten{{\underline{F}}}^{c}) + \mathcal{O}(\|\ten{\underline{F}}^{g}-\ten{{\underline{F}}}^{c}\|^2)\nonumber\\
    & \approx \vec{\underline{P}}(\ten{{\underline{F}}}^{c})-\ten{\mathbb{\underline{A}}}(\ten{{\underline{F}}}^{c})\ten{\underline{F}}^{c}+ \ten{\mathbb{\underline{A}}}(\ten{{\underline{F}}}^{c})\ten{{\underline{F}}}^{g} \nonumber\\
    & \approx \vec{\underline{P}}(\ten{{\underline{F}}}^{c})-\ten{\mathbb{\underline{A}}}(\ten{{\underline{F}}}^{c})\ten{\underline{F}}^{c}+ \ten{\mathbb{\underline{A}}}(\ten{{\underline{F}}}^{c})(\vec{\tilde{\underline{F}}}^{g}+ \vec{\Bar{\underline{F}}})\nonumber\\
    & \approx \vec{\underline{P}}(\ten{{\underline{F}}}^{c})-\ten{\mathbb{\underline{A}}}(\ten{{\underline{F}}}^{c})(\ten{\underline{F}}^{c}-\vec{\Bar{\underline{F}}})+ \ten{\mathbb{\underline{A}}}(\ten{{\underline{F}}}^{c})\ten{\psi}^{g} \vec{y} \nonumber\\
    & \approx \vec{\underline{P}}(\ten{{\underline{F}}}^{c})-\ten{\mathbb{\underline{A}}}(\ten{{\underline{F}}}^{c})\vec{\tilde{\underline{F}}}^{c}+ \ten{\mathbb{\underline{A}}}(\ten{{\underline{F}}}^{c})\ten{\psi}^{g} \vec{y}\,,%\label{eq:Plin}
\end{align}
making use of the available stress $\vec{\underline{P}}(\ten{{\underline{F}}}^{c})$ and stiffness $\ten{\mathbb{\underline{A}}}(\ten{{\underline{F}}}^{c})$ data at $\ten{{\underline{F}}}^{c}$.

With this linear approximation of the material model, the residual becomes
\begin{align}
    \vec{r} & = \sum_{c\in H} \sum_{g \in C} \ten{\psi}^{gT} \vec{\underline{P}}(\ten{{\underline{F}}}^{g}) V^g \nonumber \\
    & = \sum_{c\in H} ( \sum_{g \in C} \ten{\psi}^{gT} V^g (\vec{\underline{P}}(\ten{{\underline{F}}}^{c})-\ten{\mathbb{\underline{A}}}(\ten{{\underline{F}}}^{c})\vec{\tilde{\underline{F}}}^{c})  + \sum_{g \in C} \ten{\psi}^{gT} V^g \ten{\mathbb{\underline{A}}}(\ten{{\underline{F}}}^{c})\ten{\psi}^g \vec{y} ) \nonumber \\
    & = \vec{a} + \ten{B}\vec{y}\,,\label{eq:EMSLres}
\end{align}
where all high-order operations such as inner products can be preprocessed in an offline phase. In particular, we have
\begin{equation}
    \vec{a} = \sum_{c\in H} \ten{\Bar{\psi}}^c \ten{\hat{\underline{P}}}(\ten{\underline{F}}^{c})\,,\label{eq:S}
\end{equation}
with
\begin{equation}
    \ten{\Bar{\psi}}^c = \sum_{g\in C} \ten{\psi}^g V^g\,, \quad \text{and} \quad 
    \ten{\hat{\underline{P}}}(\ten{\underline{F}}^{c}) = \vec{\underline{P}}(\ten{{\underline{F}}}^{c})-\ten{\mathbb{\underline{A}}}(\ten{{\underline{F}}}^{c})\vec{\tilde{\underline{F}}}^{c}\,.\label{eq:EMSLPhat}
\end{equation}
The entries of $\ten{B}$ can be written in index notation as\footnote{{The implicit summation indices run in ranges of $\alpha,\beta\in \{1,...,d\}, \gamma,\delta \in \{1,...,9\}$. }}
\begin{equation}
    B_{\alpha\beta} = \sum_{c\in H} \sum_{g\in C} \psi_{\gamma\alpha}^g \mathbb{\underline{A}}_{\gamma\delta}(\vec{\underline{F}}^{c}) \psi_{\delta\beta}^g V^g = \sum_{c\in H} \mathbb{\underline{A}}_{\gamma\delta}(\vec{\underline{F}}^{c}) \sum_{g\in C} \psi_{\gamma\alpha}^g \psi_{\delta\beta}^g V^g = \sum_{c\in H} \mathbb{\underline{A}}_{\gamma\delta}(\vec{\underline{F}}^{c}) \mathcal{D}_{\gamma\alpha\delta\beta}^c\,,\label{eq:T}
\end{equation}
with the fourth-order tensor
\begin{equation}
    \mathcal{D}_{\gamma\alpha\delta\beta}^c = \sum_{g\in C} \psi_{\gamma\alpha}^g \psi_{\delta \beta}^g V^g\,.\label{eq:EMSLD}
\end{equation}
As can be seen from Eq.~\eqref{eq:EMSLres}, the coefficient $\ten{B}$ becomes the reduced stiffness matrix
\begin{equation}
    \ten{K} = \ten{B}\,.
\end{equation}
{In Eq.~\eqref{eq:EMSLres} we have thus obtained} an affine, $d$-dimensional problem to be solved in each load step. The material law only needs to be evaluated at $|H|$ strain values, and the projection in Eqs.~\eqref{eq:S} and~\eqref{eq:T} also only requires low-dimensional matrix multiplications. Meanwhile, the high-dimensional operations involved in the computation of {$\ten{\bar{\psi}}^c\in \mathbb{R}^{9\times d}$ and $\ten{\mathcal{D}}^c\in \mathbb{R}^{9\times d\times 9\times d}$ in Eqs.~\eqref{eq:EMSLPhat} and~\eqref{eq:EMSLD}} can be preprocessed in the offline phase. EMSL therefore cheaply and robustly allows us to compute a correction $\vec{y}$ to the predicted reduced variables $\vec{\bar{y}}$.

Using the same cluster-wise linearisation of the material behaviour, the homogenised stress becomes
\begin{align}
    \vec{\Bar{\underline{P}}} & = \frac{1}{V} \sum_{c\in H} \sum_{g\in C} \vec{\underline{P}}(\ten{\underline{F}}^{g}) V^g\nonumber \\
    & = \frac{1}{V} \sum_{c\in H} \sum_{g\in C} ( \vec{\underline{P}}(\ten{{\underline{F}}}^{c})-\ten{\mathbb{\underline{A}}}(\ten{{\underline{F}}}^{c})\vec{\tilde{\underline{F}}}^{c}+ \ten{\mathbb{\underline{A}}}(\ten{{\underline{F}}}^{c})\ten{\psi}^g \vec{y} ) V^g\nonumber \\
    & = \frac{1}{V} \sum_{c\in H} ( \ten{\hat{\underline{P}}}(\ten{\underline{F}}^{c}) \xi^c+ \ten{\mathbb{\underline{A}}}(\ten{{\underline{F}}}^{c}) \sum_{g\in C} \ten{\psi}^g V^g \vec{y} )\nonumber  \\
    & = \frac{1}{V} (\vec{c} + \ten{D}\vec{y}) \,,\label{eq:EMSLPhomog}
\end{align}
where
\begin{equation}
    \vec{c} = \sum_{c\in H} \ten{\hat{\underline{P}}}(\ten{\underline{F}}^c) \xi^c\,,\quad \text{and} \quad 
    \ten{D} = \sum_{c\in H} \ten{\mathbb{\underline{A}}}(\vec{\underline{F}}^c) \ten{\bar{\psi}}^c\,.
\end{equation}

As derived in Appendix~\ref{a:tangents}, the homogenised nominal stiffness can be approximated via
\begin{equation}
    \ten{\bar{\mathbb{\underline{A}}}} =\frac{1}{V} (\ten{\bar{\mathbb{\underline{A}}}}^{\text{Voigt}} + \ten{D}\ten{X})\,,
\end{equation}
with the Voigt estimate of the stiffness
\begin{equation}
    \ten{\bar{\mathbb{\underline{A}}}}^{\text{Voigt}} = \sum_{c\in H} \ten{\mathbb{\underline{A}}}(\vec{\underline{F}}^c) \xi^c\,.
\end{equation}
The sensitivity $\ten{X}=\frac{\partial \vec{y}}{\partial \vec{\underline{\bar{F}}}}$ is again derived (see Appendix~\ref{a:tangents}) using a perturbation of the (reduced) weak form, yielding
\begin{equation}
    \ten{K} \ten{X} = \ten{L}\,,\quad \text{where} \quad 
    \ten{L} = \sum_{c\in H} \ten{\bar{\psi}}^{cT} \ten{\mathbb{\underline{A}}}(\vec{\underline{F}}^c) = \ten{D}^T\,.
\end{equation}
This tangent is not exact, since it neglects the effect of changes in the reference stiffness $\ten{\mathbb{\underline{A}}}(\vec{\underline{F}}^c)$. We discuss why these contributions are comparatively small in Appendix~\ref{a:tangents}. In preliminary fully coupled FE\textsuperscript{2} computations, this approximation leads to a slight, but not dramatic, reduction in the rate of convergence. Improved convergence rates might be achieved via line searches or adaptive use of tangents computed via perturbation, but these developments are beyond the scope of the present work.

As desired, the incremental solution and homogenisation now become low-dimensional, linear operations. Costly high-dimensional operations involved in computing the matrix coefficients can be performed in the offline phase, which is summarised as pseudocode in Algorithm~\ref{alg:EMSL_offline}. This offline phase is relatively cheap in comparison with strain space ECM, which solves a high-dimensional sparse NNLS problem, and E3C, which requires many iterations of a gradient-based optimisation algorithm. The most expensive offline operation is the clustering step (which is equivalent to that involved in E3C). Contrary to ECM and as in E3C, stress snapshots on the microscale are not required.

\begin{algorithm}[h!]
\caption{EMSL offline phase}
\tcp{input snapshots of strains and parameter samples, integration point volumes, number of modes, number of material samples}
\KwIn{$\ten{\underline{F}}^s, \ten{\bar{\underline{F}}}^s, V^g,d,m$}
\tcp{Strain space POD basis}
$\ten{\psi},\ten{Y}^s \gets \text{POD}(\ten{\underline{F}}^s,d)$\;
\tcp{Linear model for inference to POD mode coefficients}
$\ten{M} = \ten{Y}^s \vec{\Bar{\underline{F}}}^{sT} (\vec{\Bar{\underline{F}}}^s\vec{\Bar{\underline{F}}}^{sT})^{-1}$\;
\tcp{Clustering of integration points}
$C \gets \text{cluster}((\ten{\psi}^g,V^g) \, \text{for} \, g\in G,m)$\;
\tcp{Cluster basis and volume}
$\xi^c\gets \sum_{g\in C} V^g, \quad c\in H$\;
$\ten{\psi}^c \gets \frac{1}{\xi^c} \sum_{g \in C} \ten{\psi}^g V^g, \quad c\in H$\;
\tcp{Volume-integrated cluster basis}
$\ten{\Bar{\psi}}^c \gets \sum_{g\in C} \ten{\psi}^g V^g, \quad c\in H$\;
\tcp{Jacobian basis terms}
$\mathcal{D}_{\gamma\alpha\delta\beta}^c \gets \sum_{g\in C} \psi_{\gamma\alpha}^g \psi_{\delta\beta}^g V^g, \quad c\in H$\;
\KwOut{$(\ten{\psi}^c, \xi^c, \ten{\Bar{\psi}}^c, \ten{\mathcal{D}}^c) \,\text{for} \, c\in H,\ten{M}$}
\label{alg:EMSL_offline}
\end{algorithm}

In each load increment, meanwhile, the expected mean cluster strain is estimated and the reference strain and stiffness are computed via the original material law. The computation of terms involved in the incremental problem is similarly expensive as the integration-point level projection in ECM and E3C. These operations, however, are only performed once per load increment. We thus address the main computational bottleneck in a hyperreduced model -- the number of evaluations of the material law -- not by a further reduction in integration points, but by a reduction of the number of times values at these integration points need to be queried. Pseudocode for the online phase of EMSL can be found in Algorithm~\ref{alg:EMSL_online}.

\begin{algorithm}[h!]
\caption{EMSL online phase}
\For{$\ten{\bar{\underline{F}}}$ in loadpath}{
\tcp{Expected mean cluster strain fluctuation}
$\vec{\bar{y}} \gets \ten{M} \vec{\bar{\underline{F}}}$\;
$\ten{\tilde{\underline{F}}}^c \gets \ten{\psi}^c\vec{\bar{y}}, \quad c\in H$\;
\tcp{Reference stress and stiffness}
$\vec{\underline{P}}^c, \ten{\mathbb{\underline{A}}}^c \gets \text{mat}(\ten{\underline{F}}^c), c \in H$\;
\tcp{Assembly terms}
$\ten{\hat{\underline{P}}}^c = \vec{\underline{P}}^c-\ten{\mathbb{\underline{A}}}^c\vec{\tilde{\underline{F}}}^c$\;
$\vec{a} \gets \sum_{c\in H} \ten{\Bar{\psi}}^c \ten{\hat{\underline{P}}}^c$\;
$B_{\alpha\beta} \gets \sum_{c\in H} \mathbb{\underline{A}}_{\gamma\delta}^c \mathcal{D}_{\gamma\alpha\beta\delta}^c$\;
$\vec{c} \gets \sum_{c\in H}  \ten{\hat{\underline{P}}}^c \xi^c$\;
$\ten{D} \gets \sum_{c\in H} \ten{\mathbb{\underline{A}}}^c \ten{\bar{\psi}}^c$\;

\tcp{Solve for increment}
$\Delta \vec{y} \gets \text{solve}(\vec{a} + \ten{B}\vec{y}=\vec{0})$\;
\tcp{Increment}
${\Vec{y}} \gets {\Vec{y}} + \Delta {\Vec{y}}$\; 
\tcp{Homogenise stress}
$\ten{\bar{\underline{P}}}\gets \frac{1}{V}(\vec{c}+\ten{D}\vec{y})$

\tcp{Algorithmic tangent}
$\ten{\bar{\mathbb{\underline{A}}}} \gets\frac{1}{V} \sum_{c\in H} \vec{\bar{A}}(\vec{\underline{F}}^c)\xi^c$\;
$\ten{X} \gets\text{solve}(\ten{K}\ten{X}=-\ten{D}^T)$\;
$\ten{\bar{\mathbb{\underline{A}}}}\gets \ten{\bar{\mathbb{\underline{A}}}}+ \frac{1}{V} \ten{D}\ten{X}$\;
}
\label{alg:EMSL_online}
\end{algorithm}

Contrary to E3C and ECM, and inspired by structure-preserving incremental approaches like NTFA, pRBMOR, and SCCA, we thus choose not to approximate the integration of an exact material law, but exactly integrate an approximate material law. {Fig.~\ref{fig:EMSL} provides a graphical illustration of this approach, which may be contrasted with Fig.~\ref{fig:RC}.} Contrary to both competing families of approaches, the approximation of the material law is performed around an empirical estimate of the solution (which, if known exactly, would be the optimal point for linearisation). This choice is motivated by the hypothesis that the error made in the approximation of the material law is compensated by more accurate (and potentially robust) integration, as well as by the reduced computational time resulting from the absence of Newton iterations.

\section{{Efficient reduced homogenisation of a porous hyperelastic microstructure}}\label{s:application}

\subsection{Problem setup}\label{s:setup}

\begin{figure}
    \centering
    \def\svgwidth{\linewidth}
    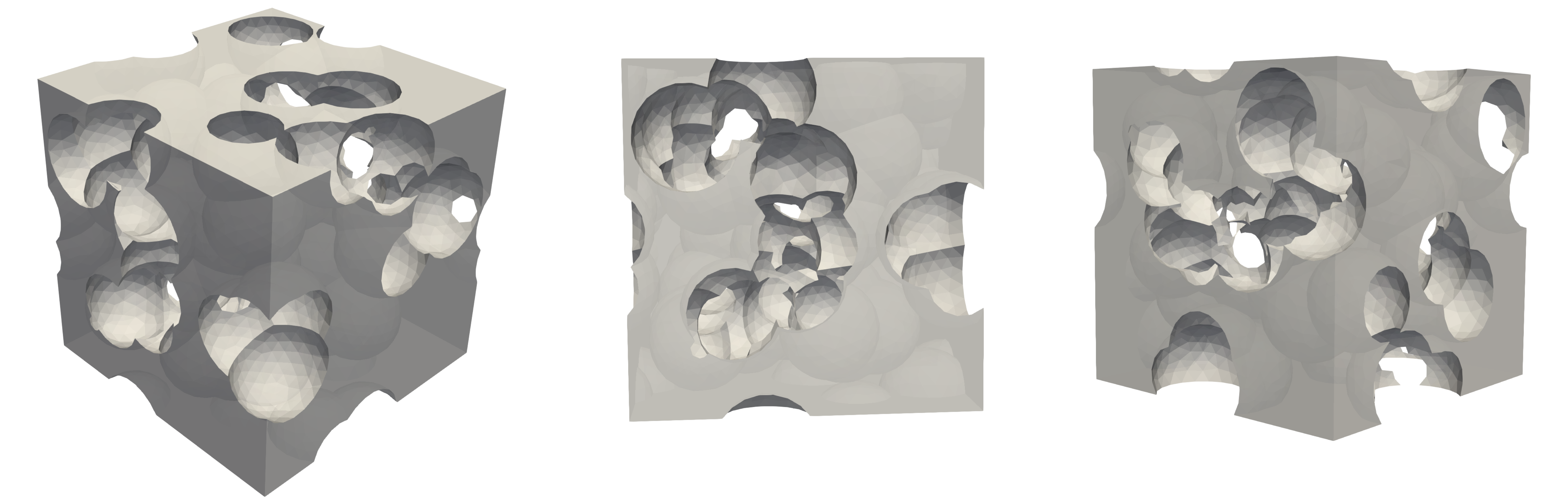
    %\includesvg[width=\linewidth]{RVE.svg}
    \caption{Example porous RVE viewed from three selected angles.}
    \label{fig:rve}
\end{figure}

{In the following, we compare the performance of the EMSL Ansatz on an artificial hyperelastic porous microstructure to that achieved with strain-space ECM and E3C. The RVE, which is shown in Fig.~\ref{fig:rve} and has previously been used in~\cite{FauSch:2026:NMO}, features 70 spherical, overlapping pores.}
The matrix is modeled via a compressible neo-Hooke material model with $E=1000, \nu = 0.25$, while the pores are treated as traction-free voids. The microstructure is discretised into 36,929 elements and 10,364 nodes using gmsh~\cite{GeuRem:2009:g3e}, resulting in 31,092 independent degrees of freedom. The relatively small size of the discretised problem means that assembly costs are a significant contributor to runtime. Problem parameters are summarised in Tab.~\ref{tab:general_params}.

\begin{table}[h!]
    \centering
    \begin{tabular}{ c c c }
        \hline
        parameter & variable & value  \\
        \hline
        Young's modulus matrix & $E$ & 1000 Nmm\textsuperscript{-2} \\
        Poisson's ratio & $\nu$ & 0.25 \\
        RVE edge length & & 2 mm \\
        number of pores & & 70 \\
        pore radii & & 0.667 mm \\
        snapshot number & $s$ & 160 \\
        validation set size & & 320 \\
        %random seed & & 42 \\
        step size & $\Delta {H}_{\text{LP}}$ & 0.025 \\
        perturbation size & $\Delta {H}_{\text{LS}}$ & 0.015 \\
        independent DOFs & $D$ & 31,092  \\ 
        element number & $n_\text{elem}$ & 36,929\\
        node number & $n_\text{node}$ & 10,364 \\
        \hline
    \end{tabular}    
    \caption{Problem and simulation parameters for the numerical experiments outlined below.}
    \label{tab:general_params}
\end{table}

\begin{figure}[h]
    \centering
    \def\svgwidth{\linewidth}
    %% Creator: Inkscape 1.2.2 (b0a8486541, 2022-12-01), www.inkscape.org
%% Accompanies image file 'RVE_defo_new.pdf' (pdf, eps, ps)
\begingroup%
  \makeatletter%
  \providecommand\color[2][]{%
    \errmessage{(Inkscape) Color is used for the text in Inkscape, but the package 'color.sty' is not loaded}%
    \renewcommand\color[2][]{}%
  }%
  \providecommand\transparent[1]{%
    \errmessage{(Inkscape) Transparency is used (non-zero) for the text in Inkscape, but the package 'transparent.sty' is not loaded}%
    \renewcommand\transparent[1]{}%
  }%
  \providecommand\rotatebox[2]{#2}%
  \newcommand*\fsize{\dimexpr\f@size pt\relax}%
  \newcommand*\lineheight[1]{\fontsize{\fsize}{#1\fsize}\selectfont}%
  \ifx\svgwidth\undefined%
    \setlength{\unitlength}{3692.16833496bp}%
    \ifx\svgscale\undefined%
      \relax%
    \else%
      \setlength{\unitlength}{\unitlength * \real{\svgscale}}%
    \fi%
  \else%
    \setlength{\unitlength}{\svgwidth}%
  \fi%
  \global\let\svgwidth\undefined%
  \global\let\svgscale\undefined%
  \makeatother%
  \begin{picture}(1,0.5439815)%
    \lineheight{1}%
    \setlength\tabcolsep{0pt}%
    \put(0,0){\includegraphics[width=\unitlength,page=1]{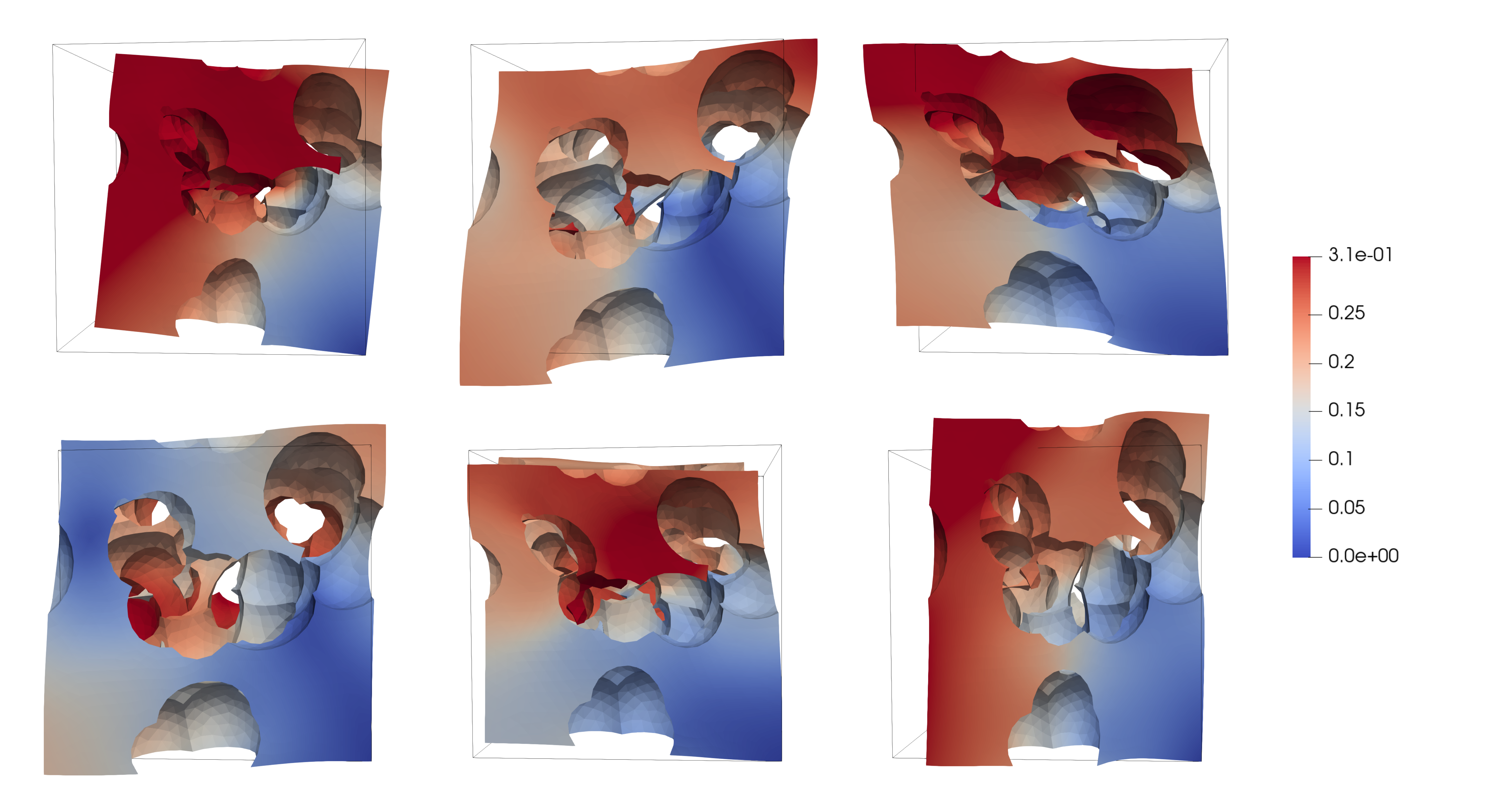}}%
    \put(0.92399699,0.26168811){\color[rgb]{0.03529412,0.03529412,0.03529412}\makebox(0,0)[lt]{\lineheight{1.25}\smash{\begin{tabular}[t]{l}$\|\bm{u}\|$\end{tabular}}}}%
  \end{picture}%
\endgroup%

    %\includesvg[width=\linewidth]{RVE_defo_new.svg}
    \caption{RVE deformation at the end of six example load paths, shown to scale. The displacement-based shading only serves to increase visual contrast and make deformations easier to see.}
    \label{fig:defos}
\end{figure}

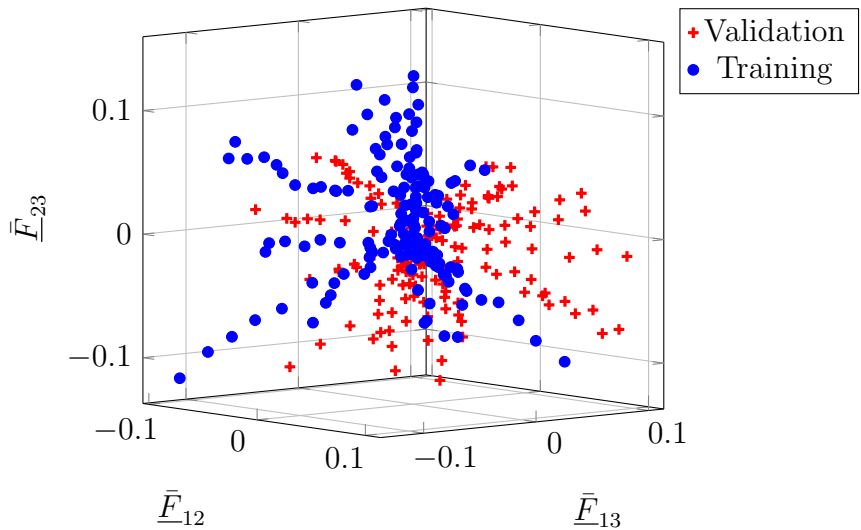
\begin{figure}[h]
\centering
\begin{tikzpicture}
    \begin{axis}
        [view={50}{20},
        xlabel=$\underline{\bar{F}}_{12}$,
        ylabel=$\underline{\bar{F}}_{13}$,
        zlabel=$\underline{\bar{F}}_{23}$,
        legend pos=outer north east,
        plot box ratio = 1 1 3,
        zmajorgrids=true,
        ymajorgrids=true,
        xmajorgrids=true,
        ]

        \addplot3[
        %scatter,
        red,
        only marks,
        very thick,
        mark=+
        ]
        table {val_sample.txt};
        \addlegendentry{Validation}

        \addplot3[
        %scatter,
        blue,
        only marks,
        mark=*
        ]
        table {train_sample.txt};
        \addlegendentry{Training}

    \end{axis}
\end{tikzpicture}
\caption{Training (blue circles) and validation (red plus signs) samples in macroscopic deformation gradient space.}
\label{fig:samples}
\end{figure}

To mimic its use in an FE\textsuperscript{2} setting, we subject the RVE to load paths with 8 load steps each. All load paths start in the undeformed configuration. Each load increment consists of two components: a step with length $\Delta \underline{F}_{\text{LP}}=0.025$ along the randomly sampled direction $\ten{N}_{\text{LP}}$ which stays constant along the load path, and a step with length $\Delta \underline{F}_{\text{LS}}=0.015$ along the direction $\ten{N}_{\text{LS}}$ which is randomly sampled in each step, such that
\begin{align}
    &\Delta \vec{\bar{\underline{F}}} = \Delta \underline{F}_{\text{LP}} \ten{N}_{\text{LP}} + \Delta \underline{F}_{\text{LS}} \ten{N}_{\text{LS}}\,, \nonumber\\&  \text{and} \quad
    \ten{\bar{\underline{F}}} \leftarrow \ten{\bar{\underline{F}}} + \Delta\ten{\bar{\underline{F}}}\,.
\end{align}
This generally leads to applied macroscopic strains of around $20\%$ at the end of each load path. For more detail about the load path definition, please consult~\cite{FauSch:2026:NMO}.
Full FEM solutions along the 8 load steps of 20 such load paths are used as snapshots $\ten{\underline{F}}^s\in \mathbb{R}^{9|G|\times s}, s=160$, on which the ECM, E3C, and EMSL models are trained. For ECM, we additionally provide the corresponding Gauss point stresses $\ten{\underline{P}}^s\in \mathbb{R}^{9|G|\times s}$ as training data. To test the reduced models, 20 further load paths are used, along the original 20. {Deformations obtained at the end of six example load paths are shown to scale in Fig.~\ref{fig:defos}. Three entries of these training and testing samples in macroscopic deformation gradient space are shown in Fig.~\ref{fig:samples}. }

The {ECM, E3C, and EMSL} models are then used to compute solutions along these testing load paths. We compute the mean relative error
\begin{equation}
    \text{Err}_\text{mean} = \frac{1}{320} \sum_{i=1}^{320} \frac{\|\Vec{\bar{\underline{P}}}^\text{i,ROM}-\vec{\bar{\underline{P}}}^\text{i,VAL}\|}{\|\Vec{\bar{\underline{P}}}^\text{i,VAL}\|}\,,\label{eq:error}
\end{equation}
in the homogenised stresses obtained from the reduced model $\Vec{\bar{\underline{P}}}^\text{i,ROM}$ relative to the full FE solution $\vec{\bar{\underline{P}}}^\text{i,VAL}$, over all samples $i \in \{1,2,..,320\}$. We also measure the online runtime required to perform these simulations\footnote{We subtract the runtime expended on the load path setup, i.e. parameter sampling, which is common to all methods, from these runtimes.}. Furthermore, parameter studies are conducted for ECM, E3C, and EMSL to measure the efficacy of these methods with different numbers $d\in [9,50]$ of Galerkin basis functions and different numbers of reduced integration or reference points $m\in [1,250]$. 
This way, we can compare the performance of the different reduced modeling approaches obtainable using a range of parameter choices in terms of the tradeoff between error and runtime. 
Since combinations of low $d$ and very high $m$ are comparatively uncompetitive in terms of this tradeoff, we did not explore such combinations in detail.% (see Tables~\ref{tab:ECSW_error},~\ref{tab:E3C_error}, and~\ref{tab:EMSL_error}). 

For ease of reading, we refer to $m$ as the number of integration points in the following. Of course, these should more properly be denoted as material reference points in the case of EMSL. Because this would make descriptions unnecessarily lengthy, we instead use consistent, if slightly inaccurate, terminology for all investigated methods.

All simulations are run using our in-house python FE framework. For linear solves in the high-order model, we use the UMFPACK multifrontal sparse LU solver~\cite{dav:2004:a8u}. To facilitate comparisons, all reduced models make use of the same routines and programming styles in the online phase. Only the selection of integration points, and which specific terms are accumulated during the assembly, change (please compare Algorithms~\ref{alg:ECSW_online} and~\ref{alg:EMSL_online}). Of course, runtime comparisons are always specific to the programming environment in which they are performed. We therefore do not seek to investigate absolute runtimes here, but to provide the reader with a rough estimate of the order of magnitude of the computational cost reduction achievable with the MOR techniques discussed above, and to compare these methods with each other, within the same framework.

\subsection{Results}\label{s:results}

Tabs.~\ref{tab:ECSW_error},~\ref{tab:E3C_error}, and~\ref{tab:EMSL_error} summarise the error levels, as defined in Eq.~\eqref{eq:error}, obtained via ECM, E3C, and EMSL, over the range of $d$ and $m$ investigated in this work.
If simulations for one or more values of the deformation gradient $\vec{\bar{\underline{F}}}$ failed to converge for some number of bases $d$ and integration points $m$, or if the overall mean error exceeds $100\%$, the corresponding entry in these tables is marked with a $\times$ symbol. 
{Additionally, the relative error values are colour-coded. Values above $5\%$ are highlighted in (matt) red, values below $5\%$ but above $2\%$ in purple, values below $2\%$ but above $1\%$ in violet, and values below $1\%$ in blue.
A graphical representation of parts of the same data can be found the complementary Figs.~\ref{fig:d12} and~\ref{fig:m100}. These highlight the error levels attained by the three investigated methods with varying $m$ and $d$, at fixed $d=12$ and fixed $m=100$, respectively.}
%a horizontal and a vertical slice through these Tables, at $d=12$ and $m =100$, respectively.

\begin{figure*}[h!]
\centering
\begin{minipage}[t]{.45\textwidth}
  \centering

    \centering
    \begin{tikzpicture}
    \begin{axis}[
        ylabel={mean relative error (\%)},
        xlabel={number of integration points $m$},
        xmin=1, xmax=100,
        ymin=0, ymax=5,
        xtick={0,20,40,60,80,100},
        ytick={0,1,2,3,4,5},
        legend style = {
        at={(0.5,1)},
        anchor=south,
        legend columns = -1,
        },
        ymajorgrids=true,
        xmajorgrids=true,
        grid style=dashed,
        table/col sep=semicolon,
        width=\linewidth,
    ]

    \addplot+[
        %scatter,
        black,
        %only marks,
        semithick,
        mark=o]
    table[x index=0,y index=3] 
    {alld12.csv};
    \addlegendentry{ECM}

    \addplot+[
        %scatter,
        red,
        %only marks,
        semithick,
        mark=star]
    table[x index=0,y index=2] 
    {alld12.csv};
    \addlegendentry{E3C}

    \addplot+[
        %scatter,
        blue,
        %only marks,
        semithick,
        mark=+]
    table[x index=0,y index=1] 
    {alld12.csv};
    \addlegendentry{EMSL}

    \end{axis}
    \end{tikzpicture}
    \caption{Relative error over number of integration points $m$, for $d=12$ for ECM (black circles), E3C (red stars), and EMSL (blue plus signs).}
    \label{fig:d12}

\end{minipage}%
%\hspace{0.5cm}
\hfill
\begin{minipage}[t]{.45\textwidth}
  \centering

    \centering
    \begin{tikzpicture}
    \begin{axis}[
        ylabel={mean relative error (\%)},
        xlabel={number of bases $d$},
        xmin=12, xmax=50,
        ymin=0, ymax=5,
        xtick={0,10,20,30,40,50},
        ytick={0,1,2,3,4,5},
        legend style = {
        at={(0.5,1)},
        anchor=south,
        legend columns = -1,
        },
        %legend pos=outer north east,
        ymajorgrids=true,
        xmajorgrids=true,
        grid style=dashed,
        table/col sep=semicolon,
        width=\linewidth,
    ]

    \addplot+[
        %scatter,
        black,
        %only marks,
        semithick,
        mark=o]
    table[x index=0,y index=3] 
    {allm100.csv};
    \addlegendentry{ECM}

    \addplot+[
        %scatter,
        red,
        %only marks,
        semithick,
        mark=star]
    table[x index=0,y index=2] 
    {allm100.csv};
    \addlegendentry{E3C}

    \addplot+[
        %scatter,
        blue,
        %only marks,
        semithick,
        mark=+]
    table[x index=0,y index=1] 
    {allm100.csv};
    \addlegendentry{EMSL}

    \end{axis}
    \end{tikzpicture}
    \caption{Relative error over number of POD bases $d$, for $m=100$ for ECM (black circles), E3C (red stars), and EMSL (blue plus signs).}
    \label{fig:m100}

\end{minipage}
\end{figure*}

\begin{table}[h!]
    \centering

\begin{adjustbox}{width=\textwidth}
    \begin{tblr}{
    colspec={cc|ccccccccccccc},
cell{3}{5} = {red!30!white},
cell{3}{6} = {red!30!white},
cell{3}{7} = {red!30!white},
cell{3}{8} = {red!30!white},
cell{3}{9} = {red!30!white},
cell{3}{10} = {red!30!white},
cell{3}{11} = {red!30!white},
cell{3}{12} = {red!30!white},
cell{4}{6} = {red!30!white},
cell{4}{7} = {red!30!white},
cell{4}{8} = {red!30!white},
cell{4}{9} = {red!30!white},
cell{4}{10} = {red!30!white},
cell{4}{11} = {red!30!white},
cell{4}{12} = {purple!30!white},
cell{4}{13} = {purple!30!white},
cell{5}{6} = {red!30!white},
cell{5}{8} = {red!30!white},
cell{5}{9} = {red!30!white},
cell{5}{10} = {red!30!white},
cell{5}{11} = {red!30!white},
cell{5}{12} = {purple!30!white},
cell{5}{13} = {purple!30!white},
cell{5}{14} = {violet!30!white},
cell{6}{7} = {red!30!white},
cell{6}{9} = {red!30!white},
cell{6}{10} = {red!30!white},
cell{6}{11} = {red!30!white},
cell{6}{12} = {red!30!white},
cell{6}{13} = {purple!30!white},
cell{6}{14} = {purple!30!white},
cell{6}{15} = {violet!30!white},
cell{7}{10} = {red!30!white},
cell{7}{11} = {red!30!white},
cell{7}{12} = {red!30!white},
cell{7}{13} = {purple!30!white},
cell{7}{14} = {violet!30!white},
cell{7}{15} = {violet!30!white},
cell{7}{16} = {blue!30!white},
cell{8}{8} = {red!30!white},
cell{8}{9} = {red!30!white},
cell{8}{10} = {red!30!white},
cell{8}{11} = {red!30!white},
cell{8}{12} = {red!30!white},
cell{8}{14} = {purple!30!white},
cell{8}{15} = {violet!30!white},    }   
        & & \SetCell[c=13]{c} $m$ & \\
        & & 1 & 2 & 5 & 10 & 15 & 20 & 30 & 40 & 50 & 75 & 100 & 150 & 200 & 250 \\
        \hline
        \SetCell[r=5]{l} $d$ & 
        9 & $\times$ & $\times$ & 91.2 & 83.0 & 46.9 & 37.3 & 13.7 & 9.32 & 7.22 & 6.37 &  &  &  &   \\
        & 12 & $\times$ & $\times$ & $\times$ & 81.5 & 55.6 & 22.6 & 16.9 & 10.4 & 7.30 & 4.17 & 3.36 &  & &  \\
        & 15 & $\times$  & $\times$ & $\times$ & 78.2 & $\times$ & 36.8 & 19.5 & 12.7 & 7.53 & 3.55 & 2.78 & 1.58 &  &  \\
        & 20 & $\times$ & $\times$ & $\times$ & $\times$ & 70.8 & $\times$ & 24.0 & 13.8 & 10.6 & 5.35 & 3.34 & 2.35 & 1.58 &  \\
        & 30 & $\times$ & $\times$ & $\times$ & $\times$ & $\times$ & $\times$ & $\times$ & 18.4 & 12.0 & 6.54 & 3.32 & 1.71 & 1.11 & 0.850  \\
        & 50 & $\times$ & $\times$ & $\times$ & $\times$ & $\times$ & 79.0 & 43.1 & 34.3 & 20.1 & 9.00 & $\times$ & 3.26 & 1.77 & $\times$ \\
    \end{tblr}    
\end{adjustbox}
    \caption{Relative error ($\%$) obtained using ECM models with different $d$ and $m$, relative to original FE simulation. Error values are colour-coded as follows: \colorbox{blue!30!white}{$\text{Err}<1\%$} in blue, \colorbox{violet!30!white}{$1\%<\text{Err}<2\%$} in violet, \colorbox{purple!30!white}{$2\%<\text{Err}<5\%$} in purple, and \colorbox{red!30!white}{$\text{Err}>5\%$} in red.}
    \label{tab:ECSW_error}
%\end{table}

%\begin{table}[h]
    \centering
\begin{adjustbox}{width=\textwidth}
    \begin{tblr}{
    colspec={cc|ccccccccccccc},
cell{3}{5} = {purple!30!white},
cell{3}{6} = {purple!30!white},
cell{3}{7} = {purple!30!white},
cell{3}{8} = {purple!30!white},
cell{3}{9} = {purple!30!white},
cell{3}{10} = {purple!30!white},
cell{3}{11} = {purple!30!white},
cell{3}{12} = {purple!30!white},
cell{4}{6} = {purple!30!white},
cell{4}{7} = {purple!30!white},
cell{4}{8} = {violet!30!white},
cell{4}{9} = {violet!30!white},
cell{4}{10} = {violet!30!white},
cell{4}{11} = {violet!30!white},
cell{4}{12} = {violet!30!white},
cell{4}{13} = {violet!30!white},
cell{5}{6} = {purple!30!white},
cell{5}{7} = {purple!30!white},
cell{5}{8} = {violet!30!white},
cell{5}{9} = {violet!30!white},
cell{5}{10} = {violet!30!white},
cell{5}{11} = {violet!30!white},
cell{5}{12} = {violet!30!white},
cell{5}{13} = {violet!30!white},
cell{5}{14} = {violet!30!white},
cell{6}{7} = {purple!30!white},
cell{6}{8} = {violet!30!white},
cell{6}{9} = {violet!30!white},
cell{6}{10} = {violet!30!white},
cell{6}{11} = {violet!30!white},
cell{6}{12} = {violet!30!white},
cell{6}{13} = {violet!30!white},
cell{6}{14} = {violet!30!white},
cell{6}{15} = {violet!30!white},
cell{7}{9} = {violet!30!white},
cell{7}{13} = {blue!30!white},
cell{7}{15} = {blue!30!white},
cell{7}{16} = {blue!30!white},
cell{8}{13} = {blue!30!white},
cell{8}{15} = {blue!30!white},
cell{8}{16} = {blue!30!white},    }   
        & & \SetCell[c=13]{c} $m$ & \\
        & & 1 & 2 & 5 & 10 & 15 & 20 & 30 & 40 & 50 & 75 & 100 & 150 & 200 & 250 \\
        \hline
        \SetCell[r=5]{l} $d$ & 
        9 & $\times$ & $\times$ & 3.39 & 2.57 & 2.07 & 2.19 & 2.27 & 2.26 & 2.32 & 2.31 &  &  &  &   \\
        & 12 & $\times$ & $\times$ & $\times$ & 2.64 & 2.32 & 1.79 & 1.84 & 1.69 & 1.75 & 1.92 & 1.89 &  & &  \\
        & 15 & $\times$ & $\times$ & $\times$ & 2.59 & 2.11 & 1.81 & 1.69 & 1.60 & 1.51 & 1.59 & 1.57 & 1.71 &  &  \\
        & 20 & $\times$ & $\times$ & $\times$ & $\times$ & 2.14 & 1.96 & 1.52 & 1.55 & 1.32 & 1.18 & 1.15 & 1.19 & 1.35 &  \\
        & 30 & $\times$ & $\times$ & $\times$ & $\times$ & $\times$ & $\times$ & 1.85 & $\times$ & $\times$ & $\times$ & 0.865 & $\times$ & 0.691 & 0.648  \\
        & 50 & $\times$ & $\times$ & $\times$ & $\times$ & $\times$ & $\times$ & $\times$ & $\times$ & $\times$ & $\times$ & 0.847 & $\times$ & 0.695 & 0.590 \\
    \end{tblr}
\end{adjustbox}
    \caption{Relative error ($\%$) obtained using E3C models with different $d$ and $m$, relative to original FE simulation. Error values are colour-coded as follows: \colorbox{blue!30!white}{$\text{Err}<1\%$} in blue, \colorbox{violet!30!white}{$1\%<\text{Err}<2\%$} in violet, \colorbox{purple!30!white}{$2\%<\text{Err}<5\%$} in purple, and \colorbox{red!30!white}{$\text{Err}>5\%$} in red.}
    \label{tab:E3C_error}

%\begin{table}[h]
    \centering
\begin{adjustbox}{width=\textwidth}
    \begin{tblr}{
    colspec={cc|ccccccccccccc},
cell{3}{3} = {purple!30!white},
cell{3}{4} = {purple!30!white},
cell{3}{5} = {violet!30!white},
cell{3}{6} = {violet!30!white},
cell{3}{7} = {violet!30!white},
cell{3}{8} = {violet!30!white},
cell{3}{9} = {violet!30!white},
cell{3}{10} = {violet!30!white},
cell{3}{11} = {violet!30!white},
cell{3}{12} = {violet!30!white},
cell{4}{3} = {purple!30!white},
cell{4}{4} = {purple!30!white},
cell{4}{5} = {violet!30!white},
cell{4}{6} = {violet!30!white},
cell{4}{7} = {violet!30!white},
cell{4}{8} = {violet!30!white},
cell{4}{9} = {violet!30!white},
cell{4}{10} = {violet!30!white},
cell{4}{11} = {violet!30!white},
cell{4}{12} = {violet!30!white},
cell{4}{13} = {violet!30!white},
cell{5}{3} = {purple!30!white},
cell{5}{4} = {purple!30!white},
cell{5}{5} = {violet!30!white},
cell{5}{6} = {violet!30!white},
cell{5}{7} = {violet!30!white},
cell{5}{8} = {violet!30!white},
cell{5}{9} = {violet!30!white},
cell{5}{10} = {violet!30!white},
cell{5}{11} = {violet!30!white},
cell{5}{12} = {blue!30!white},
cell{5}{13} = {blue!30!white},
cell{5}{14} = {blue!30!white},
cell{6}{3} = {purple!30!white},
cell{6}{4} = {purple!30!white},
cell{6}{5} = {violet!30!white},
cell{6}{6} = {violet!30!white},
cell{6}{7} = {violet!30!white},
cell{6}{8} = {violet!30!white},
cell{6}{9} = {violet!30!white},
cell{6}{10} = {violet!30!white},
cell{6}{11} = {violet!30!white},
cell{6}{12} = {purple!30!white},
cell{6}{13} = {violet!30!white},
cell{6}{14} = {blue!30!white},
cell{6}{15} = {blue!30!white},
cell{7}{3} = {purple!30!white},
cell{7}{4} = {purple!30!white},
cell{7}{5} = {violet!30!white},
cell{7}{6} = {violet!30!white},
cell{7}{7} = {violet!30!white},
cell{7}{8} = {violet!30!white},
cell{7}{9} = {violet!30!white},
cell{7}{10} = {blue!30!white},
cell{7}{11} = {blue!30!white},
cell{7}{12} = {blue!30!white},
cell{7}{13} = {blue!30!white},
cell{7}{14} = {blue!30!white},
cell{7}{15} = {blue!30!white},
cell{7}{16} = {blue!30!white},
cell{8}{3} = {purple!30!white},
cell{8}{4} = {purple!30!white},
cell{8}{5} = {violet!30!white},
cell{8}{6} = {violet!30!white},
cell{8}{7} = {violet!30!white},
cell{8}{8} = {blue!30!white},
cell{8}{9} = {blue!30!white},
cell{8}{10} = {blue!30!white},
cell{8}{11} = {blue!30!white},
cell{8}{12} = {blue!30!white},
cell{8}{13} = {blue!30!white},
cell{8}{14} = {blue!30!white},
cell{8}{15} = {blue!30!white},
cell{8}{16} = {blue!30!white},    }   
        & & \SetCell[c=13]{c} $m$ & \\
        & & 1 & 2 & 5 & 10 & 15 & 20 & 30 & 40 & 50 & 75 & 100 & 150 & 200 & 250 \\
        \hline
        \SetCell[r=5]{l} $d$ & 
        9 & 2.69 & 2.20 & 1.85 & 1.68 & 1.54 & 1.46 & 1.38 & 1.36 & 1.34 & 1.33 &  &  &  &   \\
        & 12 & 2.64 & 2.13 & 1.89 & 1.52 & 1.44 & 1.30 & 1.21 & 1.23 & 1.15 & 1.18 & 1.17 &  & &  \\
        & 15 & 2.59 & 2.07 & 1.83 & 1.52 & 1.32 & 1.21 & 1.10 & 1.09 & 1.03 & 0.993 & 0.994 & 0.985 &  &  \\
        & 20 & 2.51 & 2.43 & 1.70 & 1.40 & 1.24 & 1.18 & 1.01 & 1.10 & 1.08 & 2.04 & 1.13 & 0.916 & 0.902 &  \\
        & 30 & 2.38 & 2.26 & 1.73 & 1.48 & 1.22 & 1.58 & 1.21 & 0.771 & 0.650 & 0.620 & 0.591 & 0.578 & 0.574 & 0.568  \\
        & 50 & 2.52 & 2.35 & 1.84 & 1.37 & 1.76 & 0.893 & 0.769 & 0.679 & 0.565 & 0.506 & 0.455 & 0.514 & 0.498 & 0.760 \\
    \end{tblr}
\end{adjustbox}    
    \caption{Relative error ($\%$) obtained using EMSL models with different $d$ and $m$, relative to original FE simulation. Error values are colour-coded as follows: \colorbox{blue!30!white}{$\text{Err}<1\%$} in blue, \colorbox{violet!30!white}{$1\%<\text{Err}<2\%$} in violet, \colorbox{purple!30!white}{$2\%<\text{Err}<5\%$} in purple, and \colorbox{red!30!white}{$\text{Err}>5\%$} in red.}
    \label{tab:EMSL_error}
\end{table}

It is immediately apparent {from Tabs.~\ref{tab:ECSW_error},~\ref{tab:E3C_error}, and~\ref{tab:EMSL_error}} that E3C and EMSL yield significantly more accurate results than ECM for comparable choices of model parameters. Because {ECM} restricts its choice of integration points to a subset of the Gauss points of the original model, it requires considerably more such integration points to achieve acceptable levels of accuracy. For example, a target error level of below $2\%$ requires $150$ integration points with ECM, but only $20$ and $5$ integration points with E3C and EMSL, respectively. {In the case of ECM, only $14$, $6$, and $1$ sampled parameter combinations yield error levels below $5\%$, $2\%$, and $1\%$, respectively. With E3C, the number of sampled parameter combinations achieving the same errors levels are $41$, $28$, and $6$, respectively.}

As Fig.~\ref{fig:d12} shows, results obtained via EMSL using a given $m$ are yet more accurate than those computed via E3C. For some number of integration points $m$, EMSL can outperform E3C by up to $40\%$ in this metric. {Tab.~\ref{tab:EMSL_error} shows that all sampled parameter combinations yield error levels of below $5\%$, $61$ combinations result in errors below $2\%$, and $21$ land below a threshold of $1\%$.}
Notably, the errors attained via E3C and EMSL each converge quickly with increasing $m$, which allows for reduced simulations to be conducted with very few integration points. 

{On the evidence of Tab.~\ref{tab:EMSL_error} and Fig.~\ref{fig:m100}, }EMSL also benefits more from an increase in $d$ than the competing methods. {ECM and E3C become} prone to overfitting in this regime, which can lead to {excessive errors or} divergence in some of the validation simulations. Diverging validation simulations in E3C and ECM occur mostly for combinations of high $d$ and low $m$ -- which is an attractive region in terms of the tradeoff between accuracy and runtime. With ECM, $28$ out of $74$ parameter combinations lead to one or more diverging simulations; in the case of E3C, $33$ such combinations do. In contrast, EMSL behaves robustly over the entire parameter range: the affine formulation precludes divergence of a nonlinear solver, and error levels never become excessive.

As a result of its increased robustness and high accuracy especially when using few integration points, a target level of accuracy is attainable using EMSL with significantly smaller $d$ and $m$ than with E3C or ECM. An error level of below $2\%$, for example, can be achieved with $d=9$, $m=5$ via EMSL, but requires $d=12$, $m=20$ with E3C, and $d=15$, $m=150$ with ECM. Similarly, an error level of below $1\%$ requires $d=15$, $m=75$ or $d=50$, $m=20$ with EMSL, $d=30$, $m=100$ with E3C, and $d=30$, $m=250$ with ECM. As the runtime of the reduced models scales with $d$ and especially with $m$, this means that EMSL can meet target levels of accuracy in significantly lower runtimes than E3C or ECM.

\begin{figure}
    \centering
    \def\svgwidth{\linewidth}
    %% Creator: Inkscape 1.2.2 (b0a8486541, 2022-12-01), www.inkscape.org
%% Accompanies image file 'stress_new.pdf' (pdf, eps, ps)
\begingroup%
  \makeatletter%
  \providecommand\color[2][]{%
    \errmessage{(Inkscape) Color is used for the text in Inkscape, but the package 'color.sty' is not loaded}%
    \renewcommand\color[2][]{}%
  }%
  \providecommand\transparent[1]{%
    \errmessage{(Inkscape) Transparency is used (non-zero) for the text in Inkscape, but the package 'transparent.sty' is not loaded}%
    \renewcommand\transparent[1]{}%
  }%
  \providecommand\rotatebox[2]{#2}%
  \newcommand*\fsize{\dimexpr\f@size pt\relax}%
  \newcommand*\lineheight[1]{\fontsize{\fsize}{#1\fsize}\selectfont}%
  \ifx\svgwidth\undefined%
    \setlength{\unitlength}{2932.8782959bp}%
    \ifx\svgscale\undefined%
      \relax%
    \else%
      \setlength{\unitlength}{\unitlength * \real{\svgscale}}%
    \fi%
  \else%
    \setlength{\unitlength}{\svgwidth}%
  \fi%
  \global\let\svgwidth\undefined%
  \global\let\svgscale\undefined%
  \makeatother%
  \begin{picture}(1,0.70359901)%
    \lineheight{1}%
    \setlength\tabcolsep{0pt}%
    \put(0,0){\includegraphics[width=\unitlength,page=1]{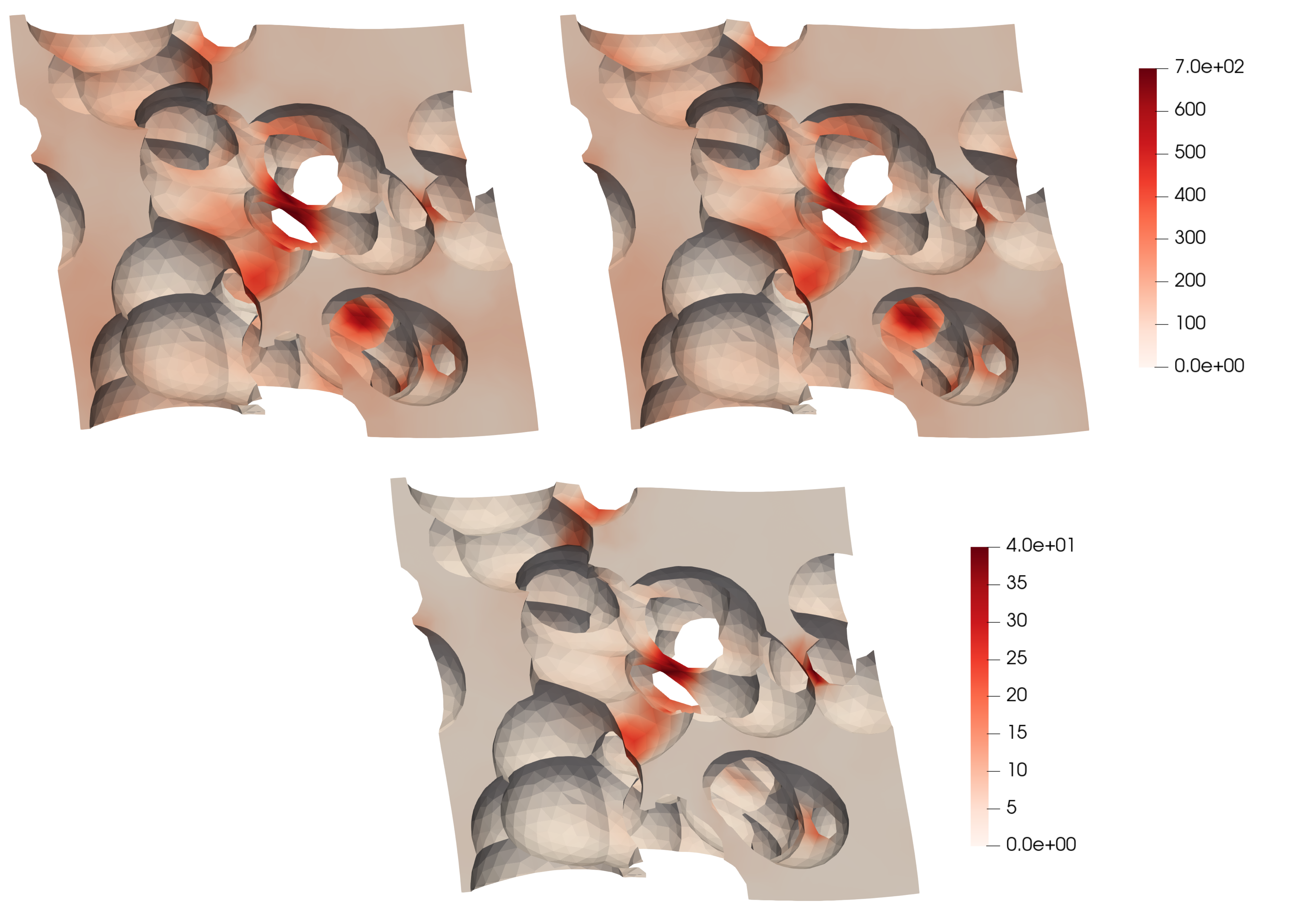}}%
    \put(0.92527653,0.52769924){\color[rgb]{0,0,0}\transparent{0.75151497}\makebox(0,0)[lt]{\lineheight{1.25}\smash{\begin{tabular}[t]{l}$\sigma^{\text{vM}}$\end{tabular}}}}%
    \put(0.79294072,0.17518586){\color[rgb]{0,0,0}\transparent{0.75151497}\makebox(0,0)[lt]{\lineheight{1.25}\smash{\begin{tabular}[t]{l}$\Delta \sigma^{\text{vM}}$\end{tabular}}}}%
  \end{picture}%
\endgroup%

    %\includesvg[width=0.8\linewidth]{stress_new.svg}
    %}
    \caption{Von Mises stress $\sigma^{\text{vM}}$ {in a slice through the RVE} at the end of one load path obtained from a full FE simulation (top left) and an EMSL simulation (top right) with $d=50,m=20$. The error $\Delta \sigma^{\text{vM}}$ is shown below. Note that the axes scales are not identical, running to $700\text{ Nmm}^{-2}$ above and $40\text{ Nmm}^{-2}$ below.}
    \label{fig:stress}
\end{figure}

While we have considered the homogenised stress so far, local quantities can also be of interest in detailed microstructure investigations. As an example, Fig.~\ref{fig:stress} compares the von Mises stresses $\sigma^{\text{vM}}$ gathered from an EMSL model with $d=50, m=20$ (top right) to those computed using a full FE simulation (top left), for an example load case. %The stress scale runs from $0$ to $700\text{ Nmm}^{-2}$.
As EMSL does not explicitly calculate stresses at every integration point during the solution procedure, these are instead reproduced from the reduced variables after convergence. It is apparent that the EMSL model captures all significant stress hotspots and the stress fields obtained by the full and reduced models are visually hard to distinguish. Therefore, the local stress error $\Delta \sigma^{\text{vM}}$ within the RVE is also shown below, with a stress scale running from $0$ to $40\text{ Nmm}^{-2}$. The worst-case relative error in critical stress hotspots in this load case and most others tends to be slightly in excess of $5\%$. This means that the reduced model can also be used to analyse trends in the structural behaviour of the RVE at the local level, while incurring low computational costs. If higher levels of accuracy are desired, single full FE simulations of the RVE could be initialised with the EMSL solution for a given load state. These converge in few Newton-Raphson iterations and can then be used to evaluate local quantities in detail.

Next, Fig.~\ref{fig:td12} highlights the growth in relative runtime with increasing $m$, for $d=12$. As expected, all methods scale close to linearly with $m$ because the evaluation of a material routine at $m$ strain values dominates the computational cost, while the solution of equation systems with $d=12$ is comparatively cheap. In ECM, slight deviations from the linear trend occur because the reduced Newton-Raphson solver employed by these methods sometimes requires slightly more or fewer iterations for convergence. Since it does not employ Newton-Raphson iterations, the runtime cost of EMSL is consistently around $60\%$ lower than that of E3C {and ECM}.

Meanwhile, the growth in runtime with increasing $d$ is shown in Fig.~\ref{fig:tm100}, for $m=100$. Runtime costs increase less steadily with $d$ {than with $m$} as larger-dimensional matrix multiplications need to be performed and (small) linear equation systems solved. However, the runtime of all methods follows a very similar trend, with EMSL being $40\%$ to $60\%$ faster than E3C, which is slightly faster than ECM.

\begin{figure*}[h!]
\centering
\begin{minipage}[t]{.45\textwidth}
  \centering

    \centering
    \begin{tikzpicture}[scale=0.9]
    \begin{axis}[
        ylabel={mean relative runtime (\%)},
        xlabel={number of integration points $m$},
        xmin=1, xmax=100,
        ymin=0, ymax=0.1,
        xtick={0,20,40,60,80,100},
        ytick={0,0.02,0.04,0.06,0.08,0.1},
        legend style = {
        at={(0.5,1)},
        anchor=south,
        legend columns = -1,
        },
        ymajorgrids=true,
        xmajorgrids=true,
        grid style=dashed,
        table/col sep=semicolon,
        width=\linewidth,
    ]

    \addplot+[
        %scatter,
        black,
        %only marks,
        semithick,
        mark=o]
    table[x index=0,y index=3] 
    {alltd12.csv};
    \addlegendentry{ECM}

    \addplot+[
        %scatter,
        red,
        %only marks,
        semithick,
        mark=star]
    table[x index=0,y index=2] 
    {alltd12.csv};
    \addlegendentry{E3C}

    \addplot+[
        %scatter,
        blue,
        %only marks,
        semithick,
        mark=+]
    table[x index=0,y index=1] 
    {alltd12.csv};
    \addlegendentry{EMSL}

    \end{axis}
    \end{tikzpicture}
    \caption{Relative runtime over number of integration points $m$, for $d=12$ for ECM (black circles), E3C (red stars), and EMSL (blue plus signs).}
    \label{fig:td12}

\end{minipage}%
\hspace{0.5cm}
\begin{minipage}[t]{.45\textwidth}
  \centering

    \centering
    \begin{tikzpicture}[scale=0.9]
    \begin{axis}[
        ylabel={mean relative runtime (\%)},
        xlabel={number of bases $d$},
        xmin=12, xmax=50,
        ymin=0, ymax=0.2,
        xtick={0,10,20,30,40,50},
        ytick={0,0.05,0.1,0.15,0.2},
        legend style = {
        at={(0.5,1)},
        anchor=south,
        legend columns = -1,
        },
        ymajorgrids=true,
        xmajorgrids=true,
        grid style=dashed,
        table/col sep=semicolon,
        width=\linewidth,
    ]

    \addplot+[
        %scatter,
        black,
        %only marks,
        semithick,
        mark=o]
    table[x index=0,y index=3] 
    {alltm100.csv};
    \addlegendentry{ECM}

    \addplot+[
        %scatter,
        red,
        %only marks,
        semithick,
        mark=star]
    table[x index=0,y index=2] 
    {alltm100.csv};
    \addlegendentry{E3C}

    \addplot+[
        %scatter,
        blue,
        %only marks,
        semithick,
        mark=+]
    table[x index=0,y index=1] 
    {alltm100.csv};
    \addlegendentry{EMSL}

    \end{axis}
    \end{tikzpicture}
    \caption{Relative runtime over number of POD bases $d$, for $m=100$ for ECM (black circles), E3C (red stars), and EMSL (blue plus signs).}
    \label{fig:tm100}

\end{minipage}
\end{figure*}

Finally, in Figs.~\ref{fig:et} and~\ref{fig:etlog} we display the combinations of relative error and relative runtime which can be achieved in the example RVE problem over all combinations of $d$ and $m$ via ECM, E3C, and EMSL. The relative error is displayed on the ordinate%$y$-axes
, while the runtime (relative to the runtime of the full FEM model) is displayed on the abscissa%$x$-axes
. Fig.~\ref{fig:et} uses a linear, and Fig.~\ref{fig:etlog} a logarithmic scale for the runtime values.% on the $x$-axis.

Both E3C and EMSL outperform ECM by a large margin in terms of the tradeoff between error and runtime. Furthermore, EMSL Pareto-dominates E3C, achieving results with a given accuracy in significantly reduced runtimes. EMSL can attain a 2000-fold speedup at an error level of ca. $0.5\%$, while E3C yields errors of ca. $1.5\%$ in the same runtime. A 10,000-fold speedup is achievable with ca. $1.0\%$ error in EMSL and ca. $2.0\%$ in E3C. A 100,000-fold acceleration is possible with ca. $2.0\%$ error in EMSL, while the parameter combinations which would afford such a runtime in E3C lead to divergence in some of the validation simulations. All methods appear ultimately to converge at higher runtimes, with application of EMSL therefore being most advantageous for applications with low runtime budgets.

\begin{figure*}[h!]
\centering
\begin{minipage}[t]{.45\textwidth}
  \centering

    \centering
    \begin{tikzpicture}
    \begin{axis}[
        ylabel={mean relative error (\%)},
        xlabel={relative runtime (\%)},
        xmin=0, xmax=0.3,
        ymin=0, ymax=5,
        xtick={0,0.05,0.1,0.15,0.2,0.25,0.3},
        ytick={0,1,2,3,4,5},
        legend style = {
        at={(0.5,1)},
        anchor=south,
        legend columns = -1,
        },
        ymajorgrids=true,
        xmajorgrids=true,
        grid style=dashed,
        table/col sep=semicolon,
        width=\linewidth,
    ]

    \addplot+[
        %scatter,
        black,
        only marks,
        semithick,
        mark=o]
    table[x index=6,y index=3] 
    {ECSW.csv};
    \addlegendentry{ECM}

    \addplot+[
        %scatter,
        red,
        only marks,
        semithick,
        mark=star]
    table[x index=6,y index=3] 
    {E3C.csv};
    \addlegendentry{E3C}

    \addplot+[
        %scatter,
        blue,
        only marks,
        semithick,
        mark=+]
    table[x index=6,y index=3] 
    {EMSL.csv};
    \addlegendentry{EMSL}

    \end{axis}
    \end{tikzpicture}
    \caption{Relative error over relative runtime for ECM (black circles), E3C (red stars), and EMSL (blue plus signs).}
    \label{fig:et}

\end{minipage}%
\hspace{0.5cm}
\begin{minipage}[t]{.45\textwidth}
  \centering

    \centering
    \begin{tikzpicture}
    \begin{semilogxaxis}[
        ylabel={mean relative error (\%)},
        xlabel={relative runtime (\%)},
        xmin=1e-4, xmax=1e0, domain=1e-4:1e0
        ymin=0, ymax=5,
        ytick={0,1,2,3,4,5},
        legend style = {
        at={(0.5,1)},
        anchor=south,
        legend columns = -1,
        },
        ymajorgrids=true,
        xmajorgrids=true,
        grid style=dashed,
        table/col sep=semicolon,
        width=\linewidth,
    ]

    \addplot+[
        %scatter,
        black,
        only marks,
        semithick,
        mark=o]
    table[x index=6,y index=3] 
    {ECSW.csv};
    \addlegendentry{ECM}

    \addplot+[
        %scatter,
        red,
        only marks,
        semithick,
        mark=star]
    table[x index=6,y index=3] 
    {E3C.csv};
    \addlegendentry{E3C}

    \addplot+[
        %scatter,
        blue,
        only marks,
        semithick,
        mark=+]
    table[x index=6,y index=3] 
    {EMSL.csv};
    \addlegendentry{EMSL}

    \end{semilogxaxis}
    \end{tikzpicture}
    \caption{Relative error over relative runtime for ECM (black circles), E3C (red stars), and EMSL (blue plus signs), logarithmic time-scale.}
    \label{fig:etlog}

\end{minipage}
\end{figure*}

\section{Summary and Outlook}\label{s:conclusion}

This work introduced a simple and efficient strain-space hyperreduction framework for hyperelastic RVEs, which we call Empirical Material Sampling and Linearisation, or EMSL. Similarly to E3C, EMSL groups the Gauss points of the underlying full FEM RVE model into clusters with similar strain behaviour. Unlike E3C, it predicts the expected average strain in each cluster once per load increment and uses this prediction as a material reference point.
Based on the stress and material tangent at this reference point in tandem with POD modes, EMSL then computes a linearised estimate of the stress response in the remainder of the cluster. This allows integration to be performed exactly via operations which can be preprocessed in the offline phase, and results in a reduced problem which is affine in each load step. Rather than approximately integrating an exact material law, EMSL thus exactly integrates an approximate material law.
We highlighted the links between this strategy and previous work on hyperreduction methods -- including general strategies such as E3C, ECM and EIM as well as incremental structure-preserving techniques such as NTFA, pRBMOR and SCCA -- in the Introduction.
In this work, we compared the performance of EMSL on a {complex porous hyperelastic microstructure} with that obtained using E3C and a strain space hybrid of ECSW and ECM. The error and runtime obtainable using all methods over a range of POD basis sizes $d$ and integration or reference points $m$ were explored in detail.

Both EMSL and E3C were found to outperform strain space ECM by a significant margin. Furthermore, EMSL Pareto-dominated E3C in the tradeoff between accuracy and runtime.
For example, EMSL attained a 2000-fold speedup at an error level of ca. $0.5\%$, while E3C yielded errors of ca. $1.5\%$ in the same runtime (the ECM variant yielded errors of around $4\%$). A 100,000-fold acceleration was possible with ca. $2.0\%$ error in EMSL, while the parameter combinations which would afford such a runtime in E3C led to divergence in some of the validation simulations. Additionally, EMSL behaved more robustly over the explored parameter range, yielding acceptable results for all 320 strain samples and 74 parameter combinations, while simulations for at least one strain sample diverged in 33 out of 74 parameter combinations when using E3C.

If slight errors are tolerable, E3C and EMSL RVE models seem to be able to operate using runtime budgets which are only one or two orders of magnitude larger than those required for a simple material routine. In this investigation, error levels of below $1\%$ could be be achieved with as few as $20$ to $100$ evaluations of the material law. With runtime accelerations on this scale, FE\textsuperscript{2} simulations can be conducted without incurring prohibitive computational costs.

Limitations, however, do remain. The cluster-wise linearisation of the material behaviour which we employ here might encounter issues in more strongly nonlinear problems. Furthermore, while the EMSL framework can be applied to history-dependent behaviour in principle, further work is required for a successful application: firstly, an iterative fixed-point variant of EMSL is likely to be necessary to accurately capture elasto-plastic behaviour. A predictor-corrector scheme must be deployed at the level of the reference points; further research is required to decide how to integrate this scheme into the estimate-and-linearise framework of EMSL. Additionally, improvements to the choice of basis are likely to be advantageous in the elasto-plastic context. %Cluster centroids might also not be effective in characterising strongly localised behaviour.

Moreover, the algorithmic tangent derived in this work neglects contributions due to changes in the cluster reference stiffness. In preliminary FE\textsuperscript{2} simulations, this led to a slight reduction in the rate of convergence. Future work should explore the application of EMSL to coupled FE\textsuperscript{2} simulations in detail, and evaluate the performance achieved with the approximate tangent proposed here to alternative strategies. These include globalising the (quasi-)Newton solver on the macro scale using line searches, selectively computing FD tangents, moment-matching reference points, or computing the algorithmic tangent exactly. The runtime performance in the FE\textsuperscript{2} context depends linearly on the number of iterations the macroscale solver requires until convergence, meaning that EMSL loses some of its runtime advantage to E3C when deployed with a reduced convergence rate resulting from a quasi-Newton scheme. Further research should investigate which approach leads to optimal performance in terms of the tradeoff between accuracy and runtime across scales.

While strain-space MOR approaches are typically applied in the RVE context only, EMSL might also be very attractive as a reduced solver for parameterised single-scale problems in solid mechanics. Provided boundary conditions can be handled successfully, the significantly reduced number of required queries to a material routine promises significant accelerations in runtime. Developing a methodology to handle Dirichlet boundary conditions, which are formulated in displacement, and not strain space, is therefore a promising avenue for further research. If this obstacle can be circumvented, strain-space MOR techniques such as EMSL or E3C become extremely attractive on account of their performance, but also their non-intrusiveness: the strain and stress data required for training can be obtained from any black-box FEM solver, and neither element routines not residual training data are necessary.

% \section*{CRediT Author Statement}

% \textbf{Erik Faust:} Conceptualization, Methodology, Software, Validation, Formal analysis, Investigation, Data Curation, Writing - Original Draft.
% \textbf{Lisa Scheunemann:} Conceptualization, Writing - Review \& Editing, Supervision, Funding acquisition.
    
\section*{Acknowledgements}

We would like to thank Dong Zhao for his contributions to the implementation of E3C. A part of this work was conducted at the Chair of Applied Mechanics (LTM) at RPTU Kaiserslautern-Landau. We gratefully acknowledge the support and funding from the LTM. We also gratefully acknowledge funding from DFG-Grant TRR 375 511263698.

% === list of references
% ------- layout-datei --------------
%\bibliographystyle{plainnat}
%\bibliographystyle{plaindin}
%\bibliographystyle{plaindin_shortname2}
%\bibliographystyle{elsarticle-num-names}
% ------- bib-datei --------------
%\printbibliography
\bibliography{template_ls}

\begin{appendices}

\section{Algorithmic tangent}\label{a:tangents}

To derive the algorithmic macroscopic tangent modulus, we rewrite the macroscopic homogenised stress (see Eq.~\eqref{eq:EMSLPhomog}) in index notation
\begin{align}
    \bar{\underline{P}}_{\alpha} &= \frac{1}{V} \sum_{c\in H} ( \underline{P}_\alpha(\vec{\underline{F}}^c)\xi^c - \mathbb{\underline{A}}_{\alpha\beta}(\vec{\underline{F}}^c)\tilde{\underline{F}}_\beta^c \xi^c + \mathbb{\underline{A}}_{\alpha\beta}(\vec{\underline{F}}^c) \bar{\psi}_{\beta\gamma}^c y_\gamma )\nonumber\\
    & = \frac{1}{V} \sum_{c\in H} ( \underline{P}_\alpha(\vec{\underline{F}}^c)\xi^c - \mathbb{\underline{A}}_{\alpha\beta}(\vec{\underline{F}}^c)\bar{\psi}_{\beta\gamma}\bar{y}_\gamma + \mathbb{\underline{A}}_{\alpha\beta}(\vec{\underline{F}}^c) \bar{\psi}_{\beta\gamma}^c y_\gamma )\nonumber\\
    & = \frac{1}{V} \sum_{c\in H} ( \underline{P}_\alpha(\vec{\underline{F}}^c)\xi^c + \mathbb{\underline{A}}_{\alpha\beta}(\vec{\underline{F}}^c)\bar{\psi}_{\beta\gamma}^c (y_\gamma-\bar{y}_\gamma))\,.
\end{align}
Noting that
\begin{equation}
    \frac{\partial f(\vec{\underline{F}}^c)}{\partial \bar{\underline{F}}_\epsilon} = \frac{\partial f(\vec{\underline{F}}^c)}{\partial \underline{F}_\zeta^c} \frac{\partial \underline{F}_\zeta^c}{\partial \bar{\underline{F}}_\epsilon} = \frac{\partial f(\vec{\underline{F}}^c)}{\partial \underline{F}_\zeta^c}(\psi_{\zeta\gamma}^c M_{\gamma\epsilon}+\delta_{\zeta\epsilon})\,,
\end{equation}
we have
\begin{equation}
\frac{\partial \underline{P}_\alpha(\vec{\underline{F}}^c)}{\partial \bar{\underline{F}}_\epsilon} = \mathbb{\underline{A}}_{\alpha\zeta}(\vec{\underline{F}}^c)(\psi_{\zeta\gamma}^c M_{\gamma\epsilon}+\delta_{\zeta\epsilon})\,,
\end{equation}
as well as
\begin{equation}
\frac{\partial \mathbb{\underline{A}}_{\alpha\beta}(\vec{\underline{F}}^c)}{\partial \bar{\underline{F}}_\epsilon} = \Gamma_{\alpha\beta\zeta}(\vec{\underline{F}}^c)(\psi_{\zeta\gamma}^c M_{\gamma\epsilon}+\delta_{\zeta\epsilon})\,,
\end{equation}
with nominal material curvature
\begin{equation}
    \Gamma_{\alpha\beta\zeta} = \frac{\partial \mathbb{\underline{A}}_{\alpha\beta}}{\partial {\underline{F}}_\zeta}\,.
\end{equation}
Then, the homogenised tangent is the derivative of the homogenised stress with respect to the macroscopic deformation gradient
\begin{align}
    \bar{A}_{\alpha\epsilon} & = \frac{\partial \bar{\underline{P}}_\alpha}{\partial \bar{\underline{F}}_\epsilon} = \frac{1}{V} \sum_{c\in H} \Big ( \mathbb{\underline{A}}_{\alpha\zeta}(\vec{\underline{F}}^c)(\psi_{\zeta\gamma}^c M_{\gamma\epsilon}+\delta_{\zeta\epsilon})\xi^c + \Gamma_{\alpha\beta\zeta}(\vec{\underline{F}}^c)(\psi_{\zeta\eta}^c M_{\eta\epsilon}+\delta_{\zeta\epsilon})\bar{\psi}_{\beta\gamma}^c (y_\gamma-\bar{y}_\gamma)\nonumber\\&\hspace{3cm} + \mathbb{\underline{A}}_{\alpha\beta}(\vec{\underline{F}}^c)\bar{\psi}_{\beta\gamma}^c \big(\frac{\partial y_\gamma}{\partial \bar{\underline{F}}_\epsilon}-\frac{\partial \bar{y}_\gamma}{\partial \bar{\underline{F}}_\epsilon}\big)\Big)\nonumber\\
    & = \frac{1}{V} \sum_{c\in H} \Big ( \mathbb{\underline{A}}_{\alpha\zeta}(\vec{\underline{F}}^c)(\psi_{\zeta\gamma}^c M_{\gamma\epsilon}+\delta_{\zeta\epsilon})\xi^c + \Gamma_{\alpha\beta\zeta}(\vec{\underline{F}}^c)(\psi_{\zeta\eta}^c M_{\eta\epsilon}+\delta_{\zeta\epsilon})\bar{\psi}_{\beta\gamma}^c (y_\gamma-\bar{y}_\gamma)\nonumber\\&\hspace{3cm} + \mathbb{\underline{A}}_{\alpha\beta}(\vec{\underline{F}}^c)\bar{\psi}_{\beta\gamma}^c \big(\frac{\partial y_\gamma}{\partial \bar{\underline{F}}_\epsilon}-M_{\gamma\epsilon}\big)\Big)\nonumber\\
    & = \frac{1}{V} \sum_{c\in H} \Big ( \mathbb{\underline{A}}_{\alpha\epsilon}(\vec{\underline{F}}^c)\xi^c + \Gamma_{\alpha\beta\zeta}(\vec{\underline{F}}^c)(\psi_{\zeta\eta}^c M_{\eta\epsilon}+\delta_{\zeta\epsilon})\bar{\psi}_{\beta\gamma}^c (y_\gamma-\bar{y}_\gamma) + \mathbb{\underline{A}}_{\alpha\beta}(\vec{\underline{F}}^c)\bar{\psi}_{\beta\gamma}^c \frac{\partial y_\gamma}{\partial \bar{\underline{F}}_\epsilon}\Big)\nonumber\\
    & = \frac{1}{V} \Big ( \bar{A}_{\alpha\epsilon} + V_{\alpha\gamma} X_{\gamma\epsilon}+  Q_{\alpha\epsilon\gamma} (y_\gamma-\bar{y}_\gamma) \Big)\,,
\end{align}
with
\begin{equation}
    \bar{A}_{\alpha\epsilon} = \sum_{c\in H} \mathbb{\underline{A}}_{\alpha\epsilon}(\vec{\underline{F}}^c)\xi^c\,,\quad 
    Q_{\alpha\epsilon\gamma} = \sum_{c\in H} \Gamma_{\alpha\beta\zeta}(\vec{\underline{F}}^c) M_{\zeta\epsilon}^c \bar{\psi}_{\beta\gamma}^c\,,
\end{equation}
and
\begin{equation}
    X_{\gamma\epsilon} = \frac{\partial y_\gamma}{\partial {\bar{\underline{F}}}_\epsilon}\,,
\end{equation}
as well as
\begin{equation}
    M_{\gamma\epsilon}^c = \psi_{\zeta\gamma}^c M_{\gamma\epsilon}+\delta_{\zeta\epsilon}\,.
\end{equation}

In order to derive the sensitivity $\ten{X} = \frac{\partial \vec{y}}{\partial \vec{\bar{\underline{F}}}}$ of the reduced solution to changes in the parameters, we will perturb the discretised weak form
\begin{equation}
    0 = \delta y_\alpha \sum_{c \in H} \sum_{g\in C} \psi_{\beta \alpha}^g \underline{P}_\beta(\vec{\underline{F}}^g) V^g = \delta y_\alpha \sum_{c \in H} \sum_{g\in C} \psi_{\beta \alpha}^g \left ( \underline{P}_\beta(\vec{\underline{F}}^c) + \mathbb{\underline{A}}_{\beta\gamma}(\vec{\underline{F}}^c) (\tilde{\underline{F}}_\gamma^g - \tilde{\underline{F}}_\gamma^c ) \right ) V^g\,,
\end{equation}
to yield
\begin{align}
    0 & = \delta y_\alpha \sum_{c \in H} \sum_{g\in C} \psi_{\beta \alpha}^g \Big ( \mathbb{\underline{A}}_{\beta\gamma}(\vec{\underline{F}}^c)\Delta \underline{F}_\gamma^c + \Gamma_{\beta\gamma\delta}(\vec{\underline{F}}^c) \Delta \underline{F}_\delta^c (\tilde{\underline{F}}_\gamma^g - \tilde{\underline{F}}_\gamma^c ) \nonumber \\ & \hspace{3.5cm} + \mathbb{\underline{A}}_{\beta\gamma}(\vec{\underline{F}}^c) ( \Delta \tilde{\underline{F}}_\gamma^g - \Delta \tilde{\underline{F}}_\gamma^c ) \Big )V^g\nonumber\\
    & = \delta y_\alpha \sum_{c \in H} \sum_{g\in C} \psi_{\beta \alpha}^g \Big ( \mathbb{\underline{A}}_{\beta\gamma}(\vec{\underline{F}}^c) (\Delta \bar{\underline{F}}_\gamma+\Delta \tilde{\underline{F}}_\gamma^c) + \Gamma_{\beta\gamma\delta}(\vec{\underline{F}}^c) (\Delta \bar{\underline{F}}_\delta + \Delta \tilde{\underline{F}}_\delta^c) (\tilde{\underline{F}}_\gamma^g - \tilde{\underline{F}}_\gamma^c )  \nonumber\\ &  \hspace{3.5cm}+ \mathbb{\underline{A}}_{\beta\gamma}(\vec{\underline{F}}^c) ( \Delta \tilde{\underline{F}}_\gamma^g - \Delta \tilde{\underline{F}}_\gamma^c ) \Big )V^g \nonumber\\
    & = \delta y_\alpha \sum_{c \in H} \sum_{g\in C} \psi_{\beta \alpha}^g \left ( \mathbb{\underline{A}}_{\beta\gamma}(\vec{\underline{F}}^c) (\Delta \bar{\underline{F}}_\gamma+\Delta \tilde{\underline{F}}_\gamma^g) + \Gamma_{\beta\gamma\delta}(\vec{\underline{F}}^c) (\Delta \bar{\underline{F}}_\delta + \Delta \tilde{\underline{F}}_\delta^c) (\tilde{\underline{F}}_\gamma^g - \tilde{\underline{F}}_\gamma^c ) \right )V^g\,.
\end{align}
Substitution yields
\begin{align}
    0 &  = \delta y_\alpha \sum_{c \in H} \sum_{g\in C} \psi_{\beta \alpha}^g \Big ( \mathbb{\underline{A}}_{\beta\gamma}(\vec{\underline{F}}^c) (\Delta \bar{\underline{F}}_\gamma+ \psi_{\gamma\delta}^g \Delta y_\delta ) \nonumber \\ &\hspace{3.5cm} + \Gamma_{\beta\gamma\delta}(\vec{\underline{F}}^c) (\Delta \bar{\underline{F}}_\delta + \psi_{\delta\epsilon} M_{\epsilon\zeta} \Delta \bar{\underline{F}}_\delta) (\psi_{\gamma\eta}^g y_\eta - \psi_{\gamma\eta}^g \bar{y}_\eta ) \Big )V^g\nonumber\\
    & = \delta y_\alpha \sum_{c \in H} \Big ( \bar{\psi}_{\beta \alpha}^c \mathbb{\underline{A}}_{\beta\gamma}(\vec{\underline{F}}^c) \Delta \bar{\underline{F}}_\gamma+ \mathbb{\underline{A}}_{\beta\gamma}(\vec{\underline{F}}^c) \sum_{g\in C} \psi_{\beta \alpha}^g V^g \psi_{\gamma\delta}^g \Delta y_\delta \nonumber \\ &\hspace{3.5cm} + \Gamma_{\beta\gamma\delta}(\vec{\underline{F}}^c)  M_{\epsilon\zeta}^c \Delta \bar{\underline{F}}_\delta (\sum_{g\in C} \psi_{\beta \alpha}^g V^g\psi_{\gamma\eta}^g y_\eta - \psi_{\beta_\alpha}^c V^c \psi_{\gamma\eta}^c \bar{y}_\eta ) \Big )\nonumber\\
    & = \delta y_\alpha \sum_{c \in H} \Big ( \bar{\psi}_{\beta \alpha}^c \mathbb{\underline{A}}_{\beta\gamma}(\vec{\underline{F}}^c) \Delta \bar{\underline{F}}_\gamma+ \mathbb{\underline{A}}_{\beta\gamma}(\vec{\underline{F}}^c) D_{\beta\alpha\gamma\delta}^c \Delta y_\delta  \nonumber \\ &\hspace{3.5cm} + \Gamma_{\beta\gamma\delta}(\vec{\underline{F}}^c)  M_{\epsilon\zeta}^c \Delta \bar{\underline{F}}_\delta (D_{\beta\alpha\gamma\eta }^c y_\eta - \psi_{\beta_\alpha}^c V^c \psi_{\gamma\eta}^c \bar{y}_\eta ) \Big )\nonumber\\ 
    & = \delta y_\alpha( L_{\alpha\delta} \Delta \bar{\underline{F}}_\delta + K_{\alpha\delta} \Delta y_\delta)\,,
\end{align}
where
\begin{equation}
    L_{\alpha\delta} = \sum_{c \in H}  \bar{\psi}_{\beta \alpha}^c \mathbb{\underline{A}}_{\beta\delta}(\vec{\underline{F}}^c)  + \sum_{c \in H} \Gamma_{\beta\gamma\delta}(\vec{\underline{F}}^c)  M_{\epsilon\zeta}^c (D_{\beta\alpha\gamma\eta }^c y_\eta - \psi_{\beta_\alpha}^c V^c \psi_{\gamma\eta}^c \bar{y}_\eta )\,.
\end{equation}
Consequently
\begin{equation}
    \ten{X} = -\ten{K}^{-1} \ten{L}\,.
\end{equation}

Unfortunately, both $\ten{L}$ and $\ten{Q}$ above feature terms with the material curvature $\ten{\Gamma} = \frac{\partial \ten{\mathbb{\underline{A}}}}{\partial \vec{\underline{F}}}$. It is possible to compute these via automatic differentiation or to labouriously derive them analytically for a given material formulation. However, they complicate the formulation and contribute to computational cost. Under some conditions, however, the offending terms vanish or become comparatively small: if $\vec{y}$ and $\vec{\bar{y}}$ are sufficiently close, the $\ten{Q}$-contribution to $\frac{\partial \vec{\bar{\underline{P}}}}{\partial \vec{\bar{\underline{F}}}}$ vanishes. This can be ensured by running EMSL iteratively in a predictor-corrector scheme. Furthermore, the material curvature contribution to $\ten{L}$ then features a covariance term
\begin{equation}
    D_{\beta\alpha\gamma\eta }^c - \psi_{\beta_\alpha}^c V^c \psi_{\gamma\eta}^c\,.
\end{equation}
If sufficiently many clusters are chosen, this terms also tends to zero, as the variance of the bases within one cluster diminishes. Alternatively, we could introduce multiple moment-matched integration points within each cluster to ensure vanishing covariance even with few clusters. 

Alternatively, the algorithmic tangent could be computed numerically, or avoided altogether via quasi-Newton schemes on the macro scale. Each of these strategies has its own disadvantages. Since we do not perform coupled multiscale simulations in this work, we do not investigate the relative merits of these strategies -- considering the material curvature exactly, neglecting it entirely, introducing multiple moment-matched integration points per cluster, avoiding homogenising tangents via quasi-Newton schemes on the macroscale -- further.

Here, we neglect the contribution of the material curvature terms, effectively introducing a locally quadratic approximation of the stored energy function around each strain sample point $\vec{\underline{F}}^c$. That way, homogenisation can be performed via
\begin{equation}
    \ten{\bar{\mathbb{\underline{A}}}} =\frac{1}{V} (\ten{\bar{\mathbb{\underline{A}}}} + \ten{D}\ten{X})\,,
\end{equation}
and
\begin{equation}
    \ten{K} \ten{X} = \ten{L}\,,
\end{equation}
where
\begin{equation}
    \ten{\bar{\mathbb{\underline{A}}}} = \sum_{c\in H} \ten{\mathbb{\underline{A}}}(\vec{\underline{F}}^c) \xi^c\,,
\end{equation}
and 
\begin{equation}
    \ten{L} = \sum_{c\in H} \ten{\bar{\psi}}^{cT} \ten{\mathbb{\underline{A}}}(\vec{\underline{F}}^c) = \ten{D}^T\,.
\end{equation}
Note that this is not a simple Voigt approximation, as we do consider the dominant contribution to the softening term. Instead, we effectively neglect the effect of the change of $\ten{\mathbb{\underline{A}}}(\vec{\underline{F}}^c)$ from one RVE simulation to the next in the computation of $\ten{\bar{\mathbb{\underline{A}}}}$. In preliminary, fully coupled FE\textsuperscript{2} simulations, this reduces the rate of convergence of the Newton-Raphson scheme on the macroscale slightly, but not significantly. 
Detailed performance comparisons of this and alternative strategies in the FE\textsuperscript{2} context will be performed in a follow-up work.

%We will investigate how this impacts the convergence of a fully coupled FE\textsuperscript{2} scheme in a follow-up work. 

\end{appendices}

% %=== appendix
% \begin{appendix}
% %\include{appendixA}
% \end{appendix}

\end{document}